\documentclass[12pt]{article}
\usepackage{latexsym}
\usepackage{amsmath,amsfonts}
\usepackage{times}
\allowdisplaybreaks[4]
\catcode`\@=11

\newcount\hour
\newcount\minute
\newtoks\amorpm \hour=\time\divide\hour by 60\minute
=\time{\multiply\hour by 60 \global\advance\minute by-\hour}
\edef\standardtime{{\ifnum\hour<12 \global\amorpm={am}%
        \else\global\amorpm={pm}\advance\hour by-12 \fi
        \ifnum\hour=0 \hour=12 \fi
        \number\hour:\ifnum\minute<10
        0\fi\number\minute\the\amorpm}}
\edef\militarytime{\number\hour:\ifnum\minute<10 0\fi\number\minute}

\def\draftlabel#1{{\@bsphack\if@filesw {\let\thepage\relax
   \xdef\@gtempa{\write\@auxout{\string
      \newlabel{#1}{{\@currentlabel}{\thepage}}}}}\@gtempa
   \if@nobreak \ifvmode\nobreak\fi\fi\fi\@esphack}
        \gdef\@eqnlabel{#1}}
\def\@eqnlabel{}
\def\@vacuum{}
\def\marginnote#1{}
\def\draftmarginnote#1{\marginpar{\raggedright\scriptsize\tt#1}}
\def\draft{
        \pagestyle{plain}
        \overfullrule=2pt
        \oddsidemargin -.5truein
        \def\@oddhead{\sl \phantom{\today\quad\militarytime} \hfil
        \smash{\Large\sl DRAFT} \hfil \today\quad\militarytime}
        \let\@evenhead\@oddhead
        \let\label=\draftlabel
        \let\marginnote=\draftmarginnote
        \def\ps@empty{\let\@mkboth\@gobbletwo
        \def\@oddfoot{\hfil \smash{\Large\sl DRAFT} \hfil}
        \let\@evenfoot\@oddhead}
        \def\@eqnnum{(\theequation)\rlap{\kern\marginparsep\tt\@eqnlabel}%
        \global\let\@eqnlabel\@vacuum}  }

\newcommand{\rf}[1]{(\ref{#1})}
\renewcommand{\theequation}{\thesection.\arabic{equation}}
\renewcommand{\thefootnote}{\fnsymbol{footnote}}
\newcommand{\newsection}{   % Numeration of eqs. is automatic
\setcounter{equation}{0}\section}

\def\appendix#1{\addtocounter{section}{1}\setcounter{equation}{0}
\renewcommand{\thesection}{\Alph{section}}
\section*{Appendix \thesection\protect\indent \parbox[t]{11.15cm}{#1}}
\addcontentsline{toc}{section}{Appendix \thesection\ \ \ #1}}

\def\be{\begin{equation}}
\def\ee{\end{equation}}
\def\beq{\begin{eqnarray}}
\def\eeq{\end{eqnarray}}

\def\parline{\,\partial\kern -0.55em /\,\,}

\def\half{{\frac{1}{2}}}

\def\DD{{\cal D}}

\def\LL{{\cal L}}
\def\MM{{\cal M}}
\def\NN{{\cal N}}

\def\Dbf{{\bf D}}
\def\Jbf{{\bf J}}

\def\Lbf{{\bf L}}

\def\Kbf{{\bf K}}
\def\Pbf{{\bf P}}

\def\alphabf{{\boldsymbol{\alpha}}}
\def\mubf{{\boldsymbol{\mu}}}
\def\Pibf{{\boldsymbol{\Pi}}}
\def\partialbf{{\boldsymbol{\partial}}}

\def\ck{|c\rangle}
\def\phik{|\phi\rangle}
\def\phibr{\langle\phi|}

\def\xik{|\xi\rangle}

\def\smzero{{\scriptscriptstyle (0)}}
\def\smone{{\scriptscriptstyle (1)}}

\def\oplussm{{\scriptscriptstyle \oplus}}
\def\ominussm{{\scriptscriptstyle \ominus}}

\def\smzero{{\scriptscriptstyle (0)}}
\def\smone{{\scriptscriptstyle (1)}}

\def\oplussm{{\scriptscriptstyle \oplus}}
\def\ominussm{{\scriptscriptstyle \ominus}}

\def\smponetwo{{\scriptscriptstyle [1,2]}}

\def\Awt{\widetilde{A}}

\def\ewt{\widetilde{e}}

\def\Uwh{\widehat{U}}

\def\chib{\bar{\chi}}

\def\Lb{\bar{L}}
\def\Vb{\bar{V}}
\def\cb{\bar{c}}
\def\eb{\bar{e}}

\def\rb{\bar{r}}
\def\chib{\bar{\chi}}
\def\varthetab{\bar{\vartheta}}

\def\(A)dS{{\rm (A)dS}}
\def\pAdS{{\rm (A)dS}}

\def\BRST{{\rm BRST}}
\def\diag{{\rm diag}}

\def\FP{{\rm FP}}
\def\tr{{\rm tr}}
\def\total{{\rm total}}

\def\intm{{\rm intm}}
\def\I{{\rm I}}
\def\II{{\rm II}}

\begin{document}

%\draft

\begin{flushright}
FIAN-TD-2014-4 \qquad \qquad \ \ \ \ \  \\
arXiv: 1404.3712 V2 [hep-th] \\
\end{flushright}

\vspace{1cm}

\begin{center}

{\Large \bf Arbitrary spin conformal fields in (A)dS }

\vspace{2.5cm}

R.R. Metsaev%
\footnote{ E-mail: metsaev@lpi.ru
}

\vspace{1cm}

{\it Department of Theoretical Physics, P.N. Lebedev Physical
Institute, \\ Leninsky prospect 53,  Moscow 119991, Russia }

\vspace{3.5cm}

{\bf Abstract}

\end{center}

Totally symmetric arbitrary spin conformal fields in (A)dS space of even dimension greater than or equal to four are studied. Ordinary-derivative and gauge invariant Lagrangian formulation for such fields is obtained. Gauge symmetries are realized by using auxiliary fields and Stueckelberg fields. We demonstrate that Lagrangian of conformal field is decomposed into a sum of gauge invariant Lagrangians for massless, partial-massless, and massive fields. We obtain a mass spectrum of the partial-massless and massive fields and confirm the conjecture about the mass spectrum made in the earlier literature. In contrast to conformal fields in flat space, the kinetic terms of conformal fields in (A)dS space turn out to be  diagonal with respect to fields entering the Lagrangian. Explicit form of conformal transformation which maps conformal field in flat space to conformal field in (A)dS space is obtained. Covariant Lorentz-like and de-Donder like gauge conditions leading to  simple gauge-fixed Lagrangian of conformal fields are proposed. Using such gauge-fixed Lagrangian, which is invariant under global BRST transformations,
we explain how the partition function of conformal field is obtained in the framework of our approach.

\vspace{2.5cm}



\newpage
\renewcommand{\thefootnote}{\arabic{footnote}}
\setcounter{footnote}{0}

%%%%%%%%%%%%%%%%%%%%%%%%%%%%%%%%%%%%%%%%%%%%%%%%%%%%%%%%%%%%%%%%%%%%%%%%%%%%%%%
%%%%%%%%%%%%%%%%%%%%%%%%%%%%%%%%%%%%%%%%%%%%%%%%%%%%%%%%%%%%%%%%%%%%%%%%%%%%%%%
\newsection{\large Introduction}
%%%%%%%%%%%%%%%%%%%%%%%%%%%%%%%%%%%%%%%%%%%%%%%%%%%%%%%%%%%%%%%%%%%%%%%%%%%%%%%
%%%%%%%%%%%%%%%%%%%%%%%%%%%%%%%%%%%%%%%%%%%%%%%%%%%%%%%%%%%%%%%%%%%%%%%%%%%%%%%

In view of aesthetic features of conformal symmetries, conformal field theories have attracted considerable interest during long period of time (see e.g., Ref.\cite{Fradkin:1985am}).
One of characteristic features of conformal fields propagating in space-time of dimension greater than or equal to four is that Lagrangian formulations of most conformal fields involve higher derivatives. Often, higher-derivative kinetic terms entering Lagrangian formulations of conformal fields make the treatment of conformal field theories cumbersome. In Refs.\cite{Metsaev:2007fq,Metsaev:2007rw}, we developed ordinary-derivative Lagrangian formulation of conformal fields.
Attractive feature of the ordinary-derivative approach is that the kinetic terms entering Lagrangian formulation of conformal fields turn out to be conventional well known kinetic terms. This is to say that, for spin-0, spin-1, and spin-2 conformal fields, the kinetic terms in our approach turn out to be the respective Klein-Gordon, Maxwell, and Einstein-Hilbert kinetic terms. For the case of higher-spin conformal fields, the appropriate kinetic terms turn out to be Fronsdal kinetic terms. Appearance of the standard kinetic terms makes the treatment of the conformal fields easier and we believe that use of the ordinary-derivative approach leads to better understanding of conformal fields.

In Refs.\cite{Metsaev:2007fq,Metsaev:2007rw}, we dealt with conformal fields propagating in flat space. Although, in our approach, the kinetic terms of conformal fields turn out to be conventional two-derivative kinetic terms, unfortunately, those kinetic terms are not diagonal with respect to fields entering a field content of our ordinary-derivative Lagrangian formulation. On the other hand, in Refs\cite{Fradkin:1982xc,Fradkin:1981jc}, it was noted that, for the case of conformal graviton field in $(A)dS_4$ space, the four-derivative Weyl kinetic operator is factorized into product of two ordinary-derivative operators. One of the ordinary-derivative operators turns out to be the standard two-derivative kinetic operator for massless transverse graviton field, while the remaining ordinary-derivative operator turns out be, as noted in Refs.\cite{Deser:1983tm,Deser:1983mm}, two-derivative kinetic operator for spin-2 partial-massless field. This remarkable factorization property of the four-derivative operator for the conformal graviton field can also be realized at the level of Lagrangian formulation. Namely, in Ref.\cite{Deser:2012qg}, it was noted that, by using appropriate field redefinitions, the ordinary-derivative Lagrangian of the conformal graviton field in $(A)dS_4$ can be presented as a sum of Lagrangians for spin-2 massless field and spin-2 partial-massless field.

Recently, in Ref.\cite{Tseytlin:2013jya}, these results were considered in the context of higher-spin conformal fields.%
\footnote{ Up-to-date reviews of higher-spin field theories may be found in Refs.\cite{Bekaert:2010hw}.}
Namely, in Ref.\cite{Tseytlin:2013jya}, it was conjectured that higher-derivative kinetic operator of arbitrary spin-$s$ conformal field propagating in $(A)dS_{d+1}$ space can be factorized into product of ordinary-derivative kinetic operators of massless, partial-massless, and massive fields.%
\footnote{ Discussion of factorized form of higher-derivative actions for higher-spin fields may be found in Ref.\cite{Joung:2012qy}.}
Note that the partial-massless fields appear when $s>1$, while the massive fields appear when $d>3$. This conjecture suggests that ordinary-derivative Lagrangian of conformal field in (A)dS can be represented as a sum of ordinary-derivative Lagrangians for appropriate massless, partial-massless, and massive fields. In this paper, among other things, we confirm the conjecture in Ref.\cite{Tseytlin:2013jya}. Namely, for arbitrary spin-$s$ conformal field propagating in (A)dS,  we find the ordinary-derivative gauge invariant Lagrangian which is a sum of ordinary-derivative and gauge invariant Lagrangians for spin-$s$ massless, spin-$s$ partial-massless, and spin-$s$ massive fields. To obtain ordinary-derivative Lagrangian of conformal field in (A)dS, we start with our Lagrangian  of conformal field in flat space obtained in Ref.\cite{Metsaev:2007rw}. Applying conformal transformation to conformal field in flat space, we obtain ordinary-derivative Lagrangian  of conformal field in (A)dS.
We note also that, in contrast to conformal fields in flat space, the kinetic terms of conformal fields in (A)dS space turn out to be  diagonal with respect to fields entering a field content of our ordinary-derivative Lagrangian formulation.

This paper is organized as follows.

In Sec.\ref{scal}, we start with the simplest example of spin-0 conformal field in $(A)dS_{d+1}$, $d$-arbitrary. For this example, we briefly discuss some characteristic features of ordinary-derivative approach.

In Sec.\ref{vector}, we study the simplest example of conformal gauge field which is spin-1 conformal field in $(A)dS_6$. We demonstrate that ordinary-derivative Lagrangian of spin-1 conformal field in $(A)dS_6$ is a sum of Lagrangians for spin-1 massless field and spin-1 massive field. We note that, for the case of spin-1 conformal field, there are no partial-massless fields. For completeness, we also present our results for spin-1 conformal field in $(A)dS_{d+1}$ for arbitrary odd $d$.

In Sec.\ref{weyl}, we deal with spin-2 field. We start with the most popular example of spin-2 conformal field in $(A)dS_4$. For this case, Lagrangian is presented as a sum of gauge invariant Lagrangians for spin-2 massless field and spin-2 partial-massless field. Novelty of our discussion, as compared to the studies in earlier literature,  is that we use a formulation involving the Stueckelberg vector field. After this we proceed with discussion of other interesting example of spin-2 conformal field in $(A)dS_6$. For this case, Lagrangian is presented as a sum of gauge invariant Lagrangians for spin-2 massless, spin-2 partial-massless, and spin-2 massive fields. Also we extend our consideration to the case of spin-2 conformal field in $(A)dS_{d+1}$ for arbitrary odd $d$.

In Sec.\ref{lagraarbspin}, we discuss arbitrary spin conformal field in $(A)dS_{d+1}$, for arbitrary odd $d$. We demonstrate that ordinary-derivative and gauge invariant Lagrangian of conformal field in (A)dS can be presented as a sum of gauge invariant Lagrangians for massless, partial-massless, and massive fields. We propose de Donder-like gauge condition which considerably simplifies the Lagrangian of conformal field. Using such gauge condition, we introduce gauge-fixed Lagrangian which is invariant under global BRST transformations and  present our derivation of the partition function of conformal field obtained in Ref.\cite{Tseytlin:2013jya}.

In Sec.\ref{general}, we demonstrate how a knowledge of gauge transformation allows us to find gauge invariant Lagrangian for (A)dS field in a straightforward way.

In Sec.\ref{review}, we review ordinary-derivative approach to conformal fields in flat space.

Sec.\ref{transformation} is devoted  to the derivation of Lagrangian for conformal fields in (A)dS. For scalar conformal field, using the formulation in flat space and applying conformal transformation which maps conformal field in flat space to conformal field in (A)dS, we obtain Lagrangian of scalar conformal field in (A)dS.
For arbitrary spin conformal field, using gauge transformation rule of the conformal field in flat space and applying conformal transformation which maps the conformal field in flat space to conformal field in (A)dS we obtain a gauge transformation rule of the arbitrary spin conformal field in (A)dS. Using  then the gauge transformation rule of the arbitrary spin conformal field in (A)dS and our result in Sec.\ref{general}, we find the ordinary-derivative gauge invariant Lagrangian of arbitrary spin conformal field in (A)dS.

Our notation and conventions are collected in Appendix.

%%%%%%%%%%%%%%%%%%%%%%%%%%%%%%%%%%%%%%%%%%%%%%%%%%%%%%%%%%%%%%%%%%%%%%%%%%%%%%%
%%%%%%%%%%%%%%%%%%%%%%%%%%%%%%%%%%%%%%%%%%%%%%%%%%%%%%%%%%%%%%%%%%%%%%%%%%%%%%%
\newsection{\large Spin-0 conformal field in (A)dS}\label{scal}
%%%%%%%%%%%%%%%%%%%%%%%%%%%%%%%%%%%%%%%%%%%%%%%%%%%%%%%%%%%%%%%%%%%%%%%%%%%%%%%
%%%%%%%%%%%%%%%%%%%%%%%%%%%%%%%%%%%%%%%%%%%%%%%%%%%%%%%%%%%%%%%%%%%%%%%%%%%%%%%

In ordinary-derivative approach, spin-0 conformal field  is described by $k+1$ scalar fields
\be \label{07042014-01}
\phi_{k'}\,, \qquad k'=0,1,\ldots , k \,, \qquad k -\hbox{ arbitrary positive integer}
\ee
Fields $\phi_{k'}$ are scalar fields of the Lorentz algebra $so(d,1)$. Lagrangian we found takes the form
\beq
\label{07042014-02} \LL & = & (-\epsilon)^k \sum_{k' = 0}^k (-)^{k'}\LL_{k'} \,,
\\
\label{07042014-03} && \LL_{k'} \equiv  \half e\phi_{k'}\bigl(\DD^2 - m_{k'}^2\bigr) \phi_{k'}\,,
\\
\label{07042014-04} && m_{k'}^2 \equiv \rho \Big( \frac{d^2}{4} - (k + \half - k')^2\Bigr)\,,
\eeq
where $e=\det e_\mu^A$, $e_\mu^A$ stands for vielbein of (A)dS, while $\DD^2$ stands for the D'Alembert operator of (A)dS space. We use $\rho= \epsilon/R^2$, where $\epsilon = 1 (-1)$ for dS (AdS) and $R$ is radius of (A)dS. For notation, see Appendix. From \rf{07042014-02}, we see that Lagrangian of spin-0 conformal field is the sum of Lagrangians for scalar fields having square of mass parameters given in \rf{07042014-04}.

The following remarks are in order.

\noindent {\bf i}) From field content in \rf{07042014-01}, we see that, in our ordinary-derivative approach, the spin-0 conformal field is described by $k+1$ scalar fields and the corresponding Lagrangian involves two derivatives. We recall that, in the framework of higher-derivative approach spin-0 conformal field is described by single field,  while the corresponding Lagrangian involves $2k+2$ derivatives. For the illustration purposes, let us demonstrate how our approach is related to the standard higher-derivative approach. To this end consider a simplest case of  higher-derivative Lagrangian for spin-0 conformal field in $(A)dS_4$ with $k=1$ (see Refs. \cite{Fradkin:1982xc},\cite{Paneitz}),
\be \label{07042014-05}
\frac{1}{e} \LL = \half \phi (\DD^2)^2\phi  -\rho \phi\DD^2 \phi\,.
\ee
Introducing an auxiliary field $\phi_1$, Lagrangian \rf{07042014-05} can be represented in ordinary-derivative form as
\be \label{07042014-06}
\frac{1}{e} \LL = \sqrt{ 2|\rho| }\, \phi \DD^2 \phi_1 - \rho \phi\DD^2 \phi - |\rho| \phi_1^2\,.
\ee
Using, in place of the field $\phi$, a  new field $\phi_0$ defined by the relation
\be
\phi = \frac{\epsilon}{\sqrt{ 2|\rho| }}( \epsilon \phi_0 + \phi_1)\,,
\ee
it is easy to check that Lagrangian \rf{07042014-06} takes the form
\beq
\label{07042014-07} \LL & = &  - \epsilon \LL_0 + \epsilon \LL_1 \,,
\\
\label{07042014-08} && \frac{1}{e} \LL_0 \equiv \half \phi_0 \DD^2 \phi_0\,,
\\
\label{07042014-09} && \frac{1}{e} \LL_1 \equiv \half \phi_1 \DD^2 \phi_1 - \rho \phi_1\phi_1 \,.
\eeq
Plugging the values $k=1$ and $d=3$ in \rf{07042014-02}, we see that our ordinary-derivative Lagrangian for these particular values of $k$ and $d$ coincides with the one in \rf{07042014-07}-\rf{07042014-09}.

\noindent {\bf ii}) In the framework of higher-derivative approach, conformally invariant operator in $S^{d+1}$, $d$-arbitrary, which involves $2k+2$ derivatives, was found in Ref.\cite{Branson}. Our values for square of mass parameter in \rf{07042014-04} coincide with the ones in Ref.\cite{Branson}. We note that it is use of the field content in \rf{07042014-01} that allows us to find Lagrangian formulation in terms of the standard second-order D'Alembert operator. To our knowledge, for arbitrary $k$, ordinary-derivative Lagrangian \rf{07042014-02} has not been discussed in the earlier literature.

%%%%%%%%%%%%%%%%%%%%%%%%%%%%%%%%%%%%%%%%%%%%%%%%%%%%%%%%%%%%%%%%%%%%%%%%%%%%%%%
%%%%%%%%%%%%%%%%%%%%%%%%%%%%%%%%%%%%%%%%%%%%%%%%%%%%%%%%%%%%%%%%%%%%%%%%%%%%%%%
\newsection{ \large Spin-1 conformal field in (A)dS}\label{vector}
%%%%%%%%%%%%%%%%%%%%%%%%%%%%%%%%%%%%%%%%%%%%%%%%%%%%%%%%%%%%%%%%%%%%%%%%%%%%%%%
%%%%%%%%%%%%%%%%%%%%%%%%%%%%%%%%%%%%%%%%%%%%%%%%%%%%%%%%%%%%%%%%%%%%%%%%%%%%%%%

We now discuss a spin-1 conformal field in (A)dS. A spin-1 conformal field in $(A)dS_4$ is described by the Maxwell theory which is well-known and therefore is not considered in this paper. In $(A)dS_{d+1}$ with $d> 3$, Lagrangian of spin-1 conformal field involves higher derivatives. Ordinary-derivative Lagrangian formulation of spin-1 conformal field in $R^{d,1}$, $d>3$, was developed in Ref.\cite{Metsaev:2007fq}. Our purpose in this section is to develop a ordinary-derivative Lagrangian formulation of spin-1 conformal field in $(A)dS_{d+1}$, $d>3$. Because spin-1 conformal field in $(A)dS_6$ is the simplest example allowing us to demonstrate many characteristic features  of our ordinary-derivative approach we start our discussion with the presentation of our result for spin-1 conformal field in $(A)dS_6$.

%%%%%%%%%%%%%%%%%%%%%%%%%%%%%%%%%%%%%%%%%%%%%%%%%%%%%%%%%%%%%%%%%%%%%%%%%%%%%%%
%%%%%%%%%%%%%%%%%%%%%%%%%%%%%%%%%%%%%%%%%%%%%%%%%%%%%%%%%%%%%%%%%%%%%%%%%%%%%%%
\subsection{ Spin-1 conformal field in $(A)dS_6$}
%%%%%%%%%%%%%%%%%%%%%%%%%%%%%%%%%%%%%%%%%%%%%%%%%%%%%%%%%%%%%%%%%%%%%%%%%%%%%%%
%%%%%%%%%%%%%%%%%%%%%%%%%%%%%%%%%%%%%%%%%%%%%%%%%%%%%%%%%%%%%%%%%%%%%%%%%%%%%%%

{\bf Field content}. To discuss ordinary-derivative and gauge
invariant formulation of spin-1 conformal field in $(A)dS_6$ we use two vector fields
denoted by $\phi_0^A$, $\phi_1^A$ and one scalar field denoted by $\phi_1$,
{\small
\be
\begin{array}{cccc}
\phi_0^A & & \phi_1^A &
\\[12pt]
& \phi_1 & &
\end{array}
\ee }

The vector fields $\phi_0^A$, $\phi_1^A$ and the scalar field $\phi_1$ transform in the respective vector and scalar representations of the Lorentz algebra
$so(5,1)$. Now we are going to demonstrate that the vector field $\phi_0^A$ enters description of spin-1 massless field, while the vector field $\phi_1^A$ and the scalar field $\phi_1$ enter Stueckelberg description of spin-1 massive field. To this end we consider Lagrangian and gauge transformations.

\noindent {\bf Gauge invariant Lagrangian}. Lagrangian we found can be presented as
\beq
\label{02042014-01} \LL & = & -\epsilon \LL_0 + \epsilon \LL_1\,,
\\
\label{02042014-03} && \LL_0 \equiv \LL_0^1\,,
\\
\label{02042014-03x} &&  \LL_1 \equiv  \LL_1^1 + \epsilon \LL_1^0\,,
\eeq
where we use the notation
\beq
\label{13042014-01} &&  \frac{1}{e}\LL_0^1  =   \half \phi_0^A (\DD^2  - 5 \rho )\phi_0^A    + \half L_0 L_0\,,
\\
\label{13042014-02} && \frac{1}{e}\LL_1^1  =  \half \phi_1^A (\DD^2 - m_1^2 - 5\rho)\phi_1^A +  \half L_1 L_1\,,
\\
&& \frac{1}{e}\LL_1^0 = \half \phi_1(\DD^2 -m_1^2) \phi_1\,,
\\
&& \hspace{1cm} L_0 \equiv \DD^B \phi_0^B\,,
\\
&& \hspace{1cm} L_1 \equiv \DD^B \phi_1^B + |m_1|  \phi_1\,,
\\
\label{02042014-04} && \hspace{1cm} m_1^2 = 2\rho\,.
\eeq
From \rf{02042014-03}, we see that Lagrangian $\LL_0$ is formulated in terms of the vector field $\phi_0^A$, while the Lagrangian $\LL_1$ is formulated in terms of the vector field $\phi_1^A$ and the scalar field $\phi_1$.

\noindent {\bf Gauge transformations}. We now discuss gauge symmetries of the Lagrangian given in \rf{02042014-01}. To this end we introduce the following gauge transformation
parameters:
\be \label{02042014-02}
\xi_0\,,\quad  \xi_1\,.
\ee
The gauge transformation parameters in \rf{02042014-02} are scalar fields of the Lorentz algebra $so(5,1)$. We note the following gauge transformations:
\beq
\label{03042014-01} && \delta \phi_0^A = \DD^A \xi_0\,,
\\
\label{03042014-02}&& \delta \phi_1^A = \DD^A \xi_1\,,
\\
\label{03042014-03} && \delta \phi_1 = -\epsilon |m_1| \xi_1\,.
\eeq
The following remarks are in order.

\noindent {\bf i}) Lagrangian $\LL_0$ in \rf{02042014-03} is invariant under $\xi_0$ gauge transformations given in \rf{03042014-01}, while the Lagrangian $\LL_1$ in \rf{02042014-03x} is invariant under $\xi_1$ gauge transformations given in \rf{03042014-02},\rf{03042014-03}. This implies that the Lagrangian $\LL_0$ describes spin-1 massless field, while the Lagrangian $\LL_1$ describes spin-1 massive field having square of mass parameter $m_1^2$ given in \rf{02042014-04}.

\noindent {\bf ii}) From \rf{03042014-03}, we see that the scalar field transforms as a Stueckelberg field. In other words, the scalar field is realized as Stueckelberg field in our description of spin-1 conformal field.

\noindent {\bf iii}) Taking into account signs of the kinetic terms in \rf{02042014-01} it is clear that Lagrangian \rf{02042014-01} describes fields related to non-unitary representation of the conformal algebra.%
\footnote{ By now, arbitrary spin unitary representations of the conformal algebra that
are relevant for elementary particles are well understood (see, e.g.,
Refs.\cite{Siegel:1988gd,Metsaev:1995jp}). In our opinion, non-unitary representations of the conformal algebra deserve to be understood better.}

\noindent {\bf Summary}. Lagrangian of spin-1 conformal field in $(A)dS_6$ given in \rf{02042014-01} is a sum of Lagrangian $\LL_0$ \rf{02042014-03} which  describes dynamics of spin-1 massless field and Lagrangian $\LL_1$ \rf{02042014-03x} which  describes dynamics of spin-1 massive field.

\noindent {\bf Lorentz-like gauge}. Representation for Lagrangians in \rf{13042014-01},\rf{13042014-02} motivates us to introduce gauge condition which we refer to as Lorentz-like gauge,
\be
L_0 = 0 \,, \qquad L_1 = 0 \,, \hspace{1cm} \hbox{ Lorentz-like gauge}\,.
\ee

%%%%%%%%%%%%%%%%%%%%%%%%%%%%%%%%%%%%%%%%%%%%%%%%%%%%%%%%%%%%%%%%%%%%%%%%%%%%%%%
%%%%%%%%%%%%%%%%%%%%%%%%%%%%%%%%%%%%%%%%%%%%%%%%%%%%%%%%%%%%%%%%%%%%%%%%%%%%%%%
\subsection{ Spin-1 conformal field in $(A)dS_{d+1}$}
%%%%%%%%%%%%%%%%%%%%%%%%%%%%%%%%%%%%%%%%%%%%%%%%%%%%%%%%%%%%%%%%%%%%%%%%%%%%%%%
%%%%%%%%%%%%%%%%%%%%%%%%%%%%%%%%%%%%%%%%%%%%%%%%%%%%%%%%%%%%%%%%%%%%%%%%%%%%%%%

To discuss ordinary-derivative and gauge
invariant approach to spin-1 conformal field in $(A)dS_{d+1}$, for arbitrary odd $d\geq 5$, we use $k+1$ vector fields denoted by $\phi_{k'}^A$, and $k$ scalar fields denoted by $\phi_{k'}$,
\beq
&& \phi_{k'}^A\,, \hspace{2cm} k'=0,1,\ldots, k,
\nonumber\\[-12pt]
\label{03042014-04} && \hspace{8cm} k\equiv \frac{d-3}{2}\,.
\\[-12pt]
&& \phi_{k'}\,, \hspace{2cm} k'=1,2,\ldots, k\,,
\nonumber
\eeq
For the illustration purposes it is helpful to represent field content in \rf{03042014-04} as follows
{\small
$$ \hbox{Field content of spin-1 conformal field in $(A)dS_{d+1}$ for arbitrary odd  $d\geq 5$, \quad $k \equiv(d-3)/2$ } $$
\be
\begin{array}{ccccccccccc}
\phi_0^A & & \phi_1^A & &\phi_2^A  & \ldots & \ldots && \phi_{k-1}^A & &
\phi_k^A
\\[12pt]
& \phi_1 & & \phi_2 & & \ldots & \ldots &\phi_{k-1}
& & \phi_k &
\end{array}
\ee }
The vector fields $\phi_{k'}^A$ and the scalars fields $\phi_{k'}$ \rf{03042014-04} transform in the respective vector and scalar representations of the Lorentz algebra
$so(d,1)$. Our purpose is to demonstrate that the vector field $\phi_0^A$ enters description of spin-1 massless field, while the vector field $\phi_{k'}^A$ and the scalar field $\phi_{k'}$ enter Stueckelberg description of spin-1 massive field. To this end we consider Lagrangian and gauge transformations.

\noindent {\bf Gauge invariant Lagrangian}. Lagrangian we found can be presented as
\beq
\label{03042014-06} (-\epsilon)^k \LL  & = &  \LL_0 +  \sum_{k'=1}^k (-)^{k'} \LL_{k'}\,,
\\
\label{03042014-07} && \LL_0 \equiv \LL_0^1\,,
\\
\label{03042014-07x} && \LL_{k'} \equiv \LL_{k'}^1 + \epsilon \LL_{k'}^0\,,
\eeq
where we use the notation
\beq
\label{13042014-03} &&  \frac{1}{e}\LL_0^1  =   \half \phi_0^A (\DD^2  - \rho d)\phi_0^A    + \half L_0^2\,,
\\
\label{13042014-04} && \frac{1}{e}\LL_{k'}^1  =  \half \phi_{k'}^A (\DD^2 - m_{k'}^2 - \rho d)\phi_{k'}^A +  \half L_{k'} L_{k'} \,,
\\
&& \frac{1}{e}\LL_{k'}^0 = \half \phi_{k'}(\DD^2 -m_{k'}^2) \phi_{k'}\,,
\\
&& \hspace{1cm} L_0 \equiv \DD^B \phi_0^B\,,
\\
&& \hspace{1cm} L_{k'} \equiv \DD^B \phi_{k'}^B + |m_{k'}|  \phi_{k'}\,,
\\
\label{03042014-08x} && \hspace{1cm} m_{k'}^2 = \rho k'(d-2-k')\,, \qquad k' = 1,\ldots , k\,,\qquad k \equiv \frac{d-3}{2}\,.
\eeq

\noindent {\bf Gauge transformations}. We now discuss gauge symmetries of the Lagrangian given in \rf{03042014-06}. To this end we introduce the following gauge transformation
parameters:
\be \label{03042014-05}
\xi_{k'}\,,\qquad  k'=0,1,\ldots , k\,.
\ee
The gauge transformation parameters $\xi_{k'}$ in \rf{03042014-05} are scalar fields of the Lorentz algebra $so(d,1)$. We note the following gauge transformations:
\beq
\label{03042014-08} && \delta \phi_0^A = \DD^A \xi_0\,,
\\
\label{03042014-09} && \delta \phi_{k'}^A = \DD^A \xi_{k'}\,,
\\
\label{03042014-10} && \delta \phi_{k'} = - \epsilon |m_{k'}| \xi_{k'}\,, \qquad k'=1,\ldots,k\,.
\eeq

The following remarks are in order.

\noindent {\bf i}) Lagrangian $\LL_0$ in \rf{03042014-07} is invariant under $\xi_0$ gauge transformations given in \rf{03042014-08}, while the Lagrangian $\LL_{k'}$ in \rf{03042014-07x} is invariant under $\xi_{k'}$ gauge transformations given in \rf{03042014-09},\rf{03042014-10}. This implies that the Lagrangian $\LL_0$ describes spin-1 massless field, while the Lagrangian $\LL_{k'}$ describes spin-1 massive field having square of mass parameter $m_{k'}^2$ given in \rf{03042014-08x}.

\noindent {\bf ii}) From \rf{03042014-10}, we see that the scalar fields transform as  Stueckelberg fields. In other words, the scalar fields are realized as Stueckelberg fields in our description of spin-1 conformal field.

\noindent {\bf Summary}. Lagrangian of spin-1 conformal field in $(A)dS_{d+1}$ given in \rf{03042014-06} is a sum of Lagrangian $\LL_0$ \rf{03042014-07} which  describes dynamics of spin-1 massless field and Lagrangians $\LL_{k'}$ \rf{03042014-07x}, $k'=1,2,\ldots, k$, which  describe dynamics of spin-1  massive fields.

\noindent {\bf Lorentz-like gauge}. Representation for Lagrangians in \rf{13042014-03},\rf{13042014-04} motivates us to introduce gauge condition which we refer to as Lorentz-like gauge for spin-1 conformal field,
\be
L_0 = 0 \,, \qquad L_{k'} = 0 \,,  \qquad k'=1,\ldots, k\,, \hspace{1cm} \hbox{ Lorentz-like gauge}.
\ee

%%%%%%%%%%%%%%%%%%%%%%%%%%%%%%%%%%%%%%%%%%%%%%%%%%%%%%%%%%%%%%%%%%%%%%%%%%%%%%%
%%%%%%%%%%%%%%%%%%%%%%%%%%%%%%%%%%%%%%%%%%%%%%%%%%%%%%%%%%%%%%%%%%%%%%%%%%%%%%%
\newsection{\large Spin-2 conformal field in (A)dS }\label{weyl}
%%%%%%%%%%%%%%%%%%%%%%%%%%%%%%%%%%%%%%%%%%%%%%%%%%%%%%%%%%%%%%%%%%%%%%%%%%%%%%%
%%%%%%%%%%%%%%%%%%%%%%%%%%%%%%%%%%%%%%%%%%%%%%%%%%%%%%%%%%%%%%%%%%%%%%%%%%%%%%%

In this section, we study  a spin-2 conformal field in (A)dS. In $(A)dS_{d+1}$, $d\geq 3$, Lagrangian of spin-2 conformal field involves higher derivatives. Ordinary-derivative Lagrangian formulation of spin-2 conformal field in $R^{d,1}$, $d\geq 3$, was developed in Ref.\cite{Metsaev:2007fq}. Our purpose in this section is to develop a ordinary-derivative Lagrangian formulation of spin-2 conformal field in $(A)dS_{d+1}$, $d\geq 3$. Because spin-2 conformal fields in $(A)dS_4$ and $(A)dS_6$ are the simplest and important examples of spin-2 conformal field theories, we consider them separately below. These two cases allow us to demonstrate some other characteristic features of our ordinary-derivative approach which absent for the case of spin-1 field in Sec.\ref{vector}. Namely,  the spin-2 conformal field in $(A)dS_4$ is the simplest example involving partial-massless field, while the spin-2 conformal field in $(A)dS_6$ is the simplest example involving both the partial-massless and massive fields.

%%%%%%%%%%%%%%%%%%%%%%%%%%%%%%%%%%%%%%%%%%%%%%%%%%%%%%%%%%%%%%%%%%%%%%%%%%%%%%%
%%%%%%%%%%%%%%%%%%%%%%%%%%%%%%%%%%%%%%%%%%%%%%%%%%%%%%%%%%%%%%%%%%%%%%%%%%%%%%%
\subsection{ Spin-2 conformal field in $(A)dS_4$}
%%%%%%%%%%%%%%%%%%%%%%%%%%%%%%%%%%%%%%%%%%%%%%%%%%%%%%%%%%%%%%%%%%%%%%%%%%%%%%%
%%%%%%%%%%%%%%%%%%%%%%%%%%%%%%%%%%%%%%%%%%%%%%%%%%%%%%%%%%%%%%%%%%%%%%%%%%%%%%%

{\bf Field content}. To discuss ordinary-derivative and gauge
invariant formulation of spin-2 conformal field in $(A)dS_4$ we use two tensor fields
denoted by $\phi_0^{AB}$, $\phi_1^{AB}$ and one vector field denoted by $\phi_1^A$;
{\small
\be \label{03042014-11}
\begin{array}{cccc}
\phi_0^{AB} & & \phi_1^{AB} &
\\[12pt]
& \phi_1^A & &
\end{array}
\ee }
The fields $\phi_0^{AB}$, $\phi_1^{AB}$ and the field $\phi_1^A$ are the respective tensor and vector fields of the Lorentz algebra $so(3,1)$. The tensor fields  $\phi_0^{AB}$, $\phi_1^{AB}$  are symmetric and traceful. Now we are going to demonstrate that the tensor field $\phi_0^{AB}$ enters description of spin-2 massless field, while the tensor field $\phi_1^{AB}$ and the vector field $\phi_1^A$ enter gauge invariant Stueckelberg description of spin-2 partial-massless field. To this end we consider Lagrangian and gauge transformations.

\noindent {\bf Gauge invariant Lagrangian}. Lagrangian we found can be presented as
\beq
\label{03042014-12} \LL & = &  -\epsilon \LL_0 +   \epsilon \LL_1\,,
\\
\label{03042014-14} && \LL_0 \equiv \LL_0^2\,,
\\
\label{03042014-14x} && \LL_1 = \LL_1^2 +  \epsilon \LL_1^1\,,
\eeq
where we use the notation
\beq
\label{13042014-05} \frac{1}{e}\LL_0^{2} & \equiv & \frac{1}{4}  \phi_0^{AB}
(\DD^2 - 2\rho)\phi_0^{AB}
- \frac{1}{8} \phi_0^{AA} (\DD^2  + 2\rho)
\phi_0^{BB}  +   \half L_0^A L_0^A\,,\qquad
\\
\label{13042014-06} \frac{1}{e}\LL_1^{2} & \equiv & \frac{1}{4}  \phi_1^{AB}
(\DD^2 -m_1^2 - 2\rho)\phi_1^{AB}
- \frac{1}{8} \phi_1^{AA} (\DD^2 - m_1^2 + 2\rho)
\phi_1^{BB}  +   \half L_1^A L_1^A\,,\qquad
\\
\label{13042014-07} \frac{1}{e}\LL_1^{1} & \equiv & \half  \phi_1^A
(\DD^2 -m_1^2 + 3\rho)\phi_1^A +  \half L_1 L_1\,,
\\
&& L_0^A = \DD^B \phi_0^{AB} - \half \DD^A \phi_0^{BB}\,,
\\
&& L_1^A = \DD^B \phi_1^{AB} - \half \DD^A \phi_1^{BB} + |m_1|\phi_1^{A}\,,
\\
&& L_1 = \DD^B \phi_1^B + \frac{\epsilon}{2}|m_1| \phi_1^{BB}\,,
\\
\label{03042014-15} && m_1^2 = 2 \rho \,.
\eeq

\noindent {\bf Gauge transformations}. We now discuss gauge symmetries of the Lagrangian given in \rf{03042014-12}. To this end we introduce the following gauge transformation
parameters:
{\small
\be \label{03042014-16}
\begin{array}{cccc}
\xi_0^A & & \xi_1^A &
\\[12pt]
& \xi_1 & &
\end{array}
\ee }
The gauge transformation parameters $\xi_0^A$, $\xi_1^A$ and $\xi_1$ in \rf{03042014-16} are the respective vector and scalar fields of the Lorentz algebra $so(3,1)$. We note the following gauge transformations:
\beq
\label{03042014-17} && \delta \phi_0^{AB} = \DD^A \xi_0^B + \DD^B \xi_0^A\,,
\\
\label{03042014-18} && \delta \phi_1^{AB} = \DD^A \xi_1^B + \DD^B \xi_1^A + |m_1|\eta^{AB} \xi_1\,,
\\
\label{03042014-19} && \delta \phi_1^A = \DD^A \xi_1 - \epsilon |m_1| \xi_1^A\,.
\eeq

The following remarks are in order.

\noindent {\bf i}) Lagrangian $\LL_0$ in \rf{03042014-14} is invariant under $\xi_0^A$ gauge transformations given in \rf{03042014-17}, while the Lagrangian $\LL_1$ in \rf{03042014-14x} is invariant under $\xi_1^A$ and $\xi_1$ gauge transformations given in \rf{03042014-18},\rf{03042014-19}. This implies that the Lagrangian $\LL_0$ describes spin-2 massless field, while the Lagrangian $\LL_1$ describes spin-2 partial-massless field having square of mass parameter $m_1^2$ given in \rf{03042014-15}.

\noindent {\bf ii}) From \rf{03042014-18},\rf{03042014-19}, we see that the vector field $\phi_1^A$ and a trace of the tensor field $\phi_1^{AB}$ transform as Stueckelberg fields. In other words, just mentioned fields are realized as Stueckelberg fields in our description of the spin-2 conformal field. Gauging away the vector field we end up with the Lagrangian obtained in Ref.\cite{Deser:2012qg}.%
\footnote{ In four-dimensions, the ordinary-derivative description of the interacting conformal gravity involving the vector Stueckelberg field was obtained in Ref.\cite{Metsaev:2007fq} by using gauge approach in Ref.\cite{Kaku:1977pa}.  Discussion of uniqueness of
the interacting conformal gravity in four-dimensions may be found in
Ref.\cite{Boulanger:2001he}. Gauge invariant description of interacting massive fields via Stueckelberg fields turns out to be powerful (see e.g., Refs.\cite{Zinoviev:2009hu}-\cite{Metsaev:2012uy}). Therefore we think that use of Stueckelberg fields for the study of interacting conformal fields might be very helpful.}

\noindent {\bf Summary}. Lagrangian of spin-2 conformal field in $(A)dS_4$ given in \rf{03042014-12} is a sum of Lagrangian $\LL_0$ \rf{03042014-14} which  describes dynamics of spin-2 massless field and Lagrangian $\LL_1$ \rf{03042014-14x}, which  describes dynamics of spin-2 partial-massless field. Square of mass parameter of the spin-2 partial-massless field is given in \rf{03042014-15}.

\noindent {\bf de Donder-like gauge}. Representation for Lagrangians in \rf{13042014-05}-\rf{13042014-07} motivates us to introduce gauge condition which we refer to as de Donder-like gauge for spin-2 conformal field,
\be
L_0^A = 0 \,, \qquad L_1^A = 0 \,, \qquad  L_1=0 \,, \hspace{1cm} \hbox{ de Donder-like gauge}.
\ee

%%%%%%%%%%%%%%%%%%%%%%%%%%%%%%%%%%%%%%%%%%%%%%%%%%%%%%%%%%%%%%%%%%%%%%%%%%%%%%%
%%%%%%%%%%%%%%%%%%%%%%%%%%%%%%%%%%%%%%%%%%%%%%%%%%%%%%%%%%%%%%%%%%%%%%%%%%%%%%%
\subsection{ Spin-2 conformal field in $(A)dS_6$}
%%%%%%%%%%%%%%%%%%%%%%%%%%%%%%%%%%%%%%%%%%%%%%%%%%%%%%%%%%%%%%%%%%%%%%%%%%%%%%%
%%%%%%%%%%%%%%%%%%%%%%%%%%%%%%%%%%%%%%%%%%%%%%%%%%%%%%%%%%%%%%%%%%%%%%%%%%%%%%%

{\bf Field content}. To discuss ordinary-derivative and gauge
invariant formulation of spin-2 conformal field in $(A)dS_6$ we use three tensor fields
denoted by $\phi_0^{AB}$, $\phi_1^{AB}$, $\phi_2^{AB}$, two vector fields denoted by $\phi_1^A$, $\phi_2^A$ and one scalar field denoted by $\phi_2$,
{\small
\be \label{03042014-20}
\begin{array}{ccccc}
\phi_0^{AB} & & \phi_1^{AB} & &\phi_2^{AB}
\\[12pt]
& \phi_1^A & & \phi_2^A &
\\[12pt]
& & \phi_2 &&
\end{array}
\ee }
The tensor fields $\phi_0^{AB}$, $\phi_1^{AB}$, $\phi_2^{AB}$, the vector fields $\phi_1^A$, $\phi_2^A$, and the scalar field $\phi_2$ are the respective tensor, vector, and scalar  fields of the Lorentz algebra $so(5,1)$. The tensor fields  $\phi_0^{AB}$, $\phi_1^{AB}$, $\phi_2^{AB}$  are symmetric and traceful. Now we are going to demonstrate that the field $\phi_0^{AB}$ enters description of spin-2 massless field, the fields $\phi_1^{AB}$, $\phi_1^A$ enter gauge invariant Stueckelberg description of spin-2 partial-massless field, while the fields $\phi_2^{AB}$, $\phi_2^A$, $\phi_2$ enter gauge invariant Stueckelberg description of spin-2 massive field. To this end we consider Lagrangian and gauge transformations.

\noindent {\bf Gauge invariant Lagrangian}. Lagrangian we found can be presented as
\beq
\label{03042014-21} \LL & = &  \LL_0 - \LL_1 + \LL_2\,,
\\
\label{03042014-22} && \LL_0 \equiv  \LL_0^2\,,
\\
\label{03042014-23} && \LL_1\equiv  \LL_1^2 + \epsilon \LL_1^1\,,
\\
\label{03042014-24} && \LL_2 \equiv \LL_2^2 + \epsilon \LL_2^1 + \LL_2^0\,,
\eeq
where we use the notation
\beq
\label{13042014-08} \frac{1}{e}\LL_0^2 & \equiv & \frac{1}{4}  \phi_0^{AB}
(\DD^2 - 2\rho)\phi_0^{AB} - \frac{1}{8} \phi_0^{AA} (\DD^2  + 6\rho )
\phi_0^{BB}  +   \half L_0^A L_0^A\,,\qquad
\\
\label{13042014-09} \frac{1}{e}\LL_{k'}^2 & \equiv & \frac{1}{4}  \phi_{k'}^{AB}
(\DD^2 -m_{k'}^2 - 2\rho)\phi_{k'}^{AB}
- \frac{1}{8} \phi_{k'}^{AA} (\DD^2 - m_{k'}^2 + 6\rho)
\phi_{k'}^{BB}  +   \half L_{k'}^A L_{k'}^A\,,\qquad
\\
\label{13042014-10} \frac{1}{e}\LL_{k'}^1 & \equiv & \half  \phi_{k'}^A
(\DD^2 -m_{k'}^2 + 5\rho)\phi_{k'}^A +  \half L_{k'} L_{k'}\,, \hspace{1.8cm}  k'=1,2,
\\
 \frac{1}{e}\LL_2^0 & \equiv & \half  \phi_2 (\DD^2 -m_2^2 + 10\rho)\phi_2 \,,
\\
&& L_0^A = \DD^B \phi_0^{AB} - \half \DD^A \phi_0^{BB}\,,
\\
&& L_{k'}^A = \DD^B \phi_{k'}^{AB} - \half \DD^A \phi_{k'}^{BB} + |m_{k'}|\phi_{k'}^{A}\,, \hspace{1.5cm}  k'=1,2,
\\
&& L_1 = \DD^B \phi_1^B + \frac{\epsilon}{2}|m_1| \phi_1^{BB}\,,
\\
&& L_2 = \DD^B \phi_2^B + \frac{\epsilon}{2}|m_2| \phi_2^{BB} + f \phi_2\,,
\\
\label{03042014-25} && m_1^2 = 4 \rho \,, \qquad  m_2^2 = 6\rho\,,
\\
&& f   =  \sqrt{ 5 |\rho|}\,.
\eeq

\noindent {\bf Gauge transformations}. We now discuss gauge symmetries of the Lagrangian given in \rf{03042014-21}. To this end we introduce the following gauge transformation
parameters:
{\small
\be \label{03042014-26}
\begin{array}{ccccc}
\xi_0^A & & \xi_1^A & &\xi_2^A
\\[12pt]
& \xi_1 & & \xi_2&
\end{array}
\ee }
The gauge transformation parameters $\xi_0^A$, $\xi_1^A$, $\xi_2^A$ and $\xi_1$, $\xi_2$ in \rf{03042014-26} are the respective vector and scalar fields of the Lorentz algebra $so(5,1)$. We note the following gauge transformations:
\beq
\label{03042014-27} && \delta \phi_0^{AB} = \DD^A \xi_0^B + \DD^B \xi_0^A\,,
\\
\label{03042014-28} && \delta \phi_{k'}^{AB} = \DD^A \xi_{k'}^B + \DD^B \xi_{k'}^A + \half |m_{k'}| \eta^{AB} \xi_{k'}\,, \hspace{1cm} k'=1,2, \qquad
\\
\label{03042014-29} && \delta \phi_{k'}^A = \DD^A \xi_{k'} - \epsilon |m_{k'}| \xi_{k'}^A\,, \hspace{3cm} k'=1,2,
\\
\label{03042014-30} && \delta \phi_2 = - \epsilon f \xi_2\,.
\eeq

The following remarks are in order.

\noindent {\bf i}) Lagrangian $\LL_0$ in \rf{03042014-22} is invariant under $\xi_0^A$ gauge transformations given in \rf{03042014-27}. This implies that the Lagrangian $\LL_0$ describes spin-2 massless field.

\noindent {\bf ii}) Lagrangian $\LL_1$ in \rf{03042014-23} is invariant under $\xi_1^A$ and $\xi_1$ gauge transformations given in \rf{03042014-28}, \rf{03042014-29}. This implies that the Lagrangian $\LL_1$ describes spin-2 partial-massless field having square of mass parameter $m_1^2$ given in \rf{03042014-25}.

\noindent {\bf iii}) Lagrangian $\LL_2$ in \rf{03042014-24} is invariant under $\xi_2^A$ and $\xi_2$ gauge transformations given in \rf{03042014-28}-\rf{03042014-30}. This implies that the Lagrangian $\LL_2$ describes spin-2 massive field having square of mass parameter $m_2^2$ given in \rf{03042014-25}.

\noindent {\bf iv}) From \rf{03042014-27}-\rf{03042014-30}, we see that the scalar field, the vector fields, and trace of the tensor field $\phi_1^{AB}$ transform as Stueckelberg fields. In other words, just mentioned fields are realized as Stueckelberg fields in our description of spin-2 conformal field.%
\footnote{ For the first time, ordinary-derivative description of six-dimensional gravity involving  Stueckelberg fields was developed in Refs.\cite{Metsaev:2007fq,Metsaev:2010kp}. Recent discussion  of various aspects of conformal gravity in six-dimensions may be found in Ref.\cite{Lu:2011ks}.}

\noindent {\bf Summary}. Lagrangian of spin-2 conformal field in $(A)dS_6$ given in \rf{03042014-21} is a sum of Lagrangian $\LL_0$ \rf{03042014-22}, which  describes dynamics of spin-2 massless field, Lagrangian $\LL_1$ \rf{03042014-23}, which  describe dynamics of spin-2 partial-massless fields, and Lagrangian $\LL_2$, which describes dynamics of spin-2 massive field. Squares of mass parameter for partial-massless field, $m_1^2$, and the one for massive field, $m_2^2$, are given in \rf{03042014-25}.

\noindent {\bf de Donder-like gauge}. Representation for Lagrangians in \rf{13042014-08}-\rf{13042014-10} motivates us to introduce gauge condition which we refer to as de Donder-like gauges for spin-2 conformal field,
\be
L_{k'}^A = 0 \,, \quad k'=0,1,2; \qquad L_1 = 0 \,, \qquad  L_2=0 \,, \hspace{1cm} \hbox{ de Donder-like gauge}\,.
\ee

%%%%%%%%%%%%%%%%%%%%%%%%%%%%%%%%%%%%%%%%%%%%%%%%%%%%%%%%%%%%%%%%%%%%%%%%%%%%%%%
%%%%%%%%%%%%%%%%%%%%%%%%%%%%%%%%%%%%%%%%%%%%%%%%%%%%%%%%%%%%%%%%%%%%%%%%%%%%%%%
\subsection{ Spin-2 conformal field in $(A)dS_{d+1}$}
%%%%%%%%%%%%%%%%%%%%%%%%%%%%%%%%%%%%%%%%%%%%%%%%%%%%%%%%%%%%%%%%%%%%%%%%%%%%%%%
%%%%%%%%%%%%%%%%%%%%%%%%%%%%%%%%%%%%%%%%%%%%%%%%%%%%%%%%%%%%%%%%%%%%%%%%%%%%%%%

To discuss ordinary-derivative and gauge
invariant formulation of spin-2 conformal field in $(A)dS_{d+1}$, for arbitrary odd $d$,  we use $k+1$ tensor fields denoted by $\phi_{k'}^{AB}$, $k$ vector fields denoted by $\phi_{k'}^A$, and $k-1$ scalar fields denoted by $\phi_{k'}$,
\beq
\label{03042014-31} && \phi_{k'}^{AB}\,, \hspace{2cm} k'=0,1,\ldots, k,
\nonumber\\
&& \phi_{k'}^A\,, \hspace{2.2cm} k'=1,2,\ldots, k, \hspace{3cm} k \equiv \frac{d-1}{2}\,. \qquad
\\
&& \phi_{k'}\,, \hspace{2.2cm} k'=2,3,\ldots, k\,,
\nonumber
\eeq
For the illustration purposes it is helpful to represent field content in \rf{03042014-31} as follows

{\small
$$ \hbox{Field content of spin-2 conformal field in $(A)dS_{d+1}$  for  arbitrary odd $d\geq 5$, \ \ $k \equiv (d-1)/2$ }$$
\be
\begin{array}{cccccccccccc}
\phi_0^{AB} & & \phi_1^{AB} & &\phi_2^{AB}& &\phi_3^{AB} & \ldots & \ldots & \phi_{k-1}^{AB}& & \phi_k^{AB}
\\[12pt]
& \phi_1^A & & \phi_2^A & & \phi_3^A &  \ldots & \ldots & \phi_{k-1}^A& &\phi_k^A&
\\[12pt]
& & \phi_2& & \phi_3 & \ldots & \ldots & \phi_{k-1}& & \phi_k & &
\end{array}
\ee }

The tensor fields $\phi_{k'}^{AB}$, the vector fields $\phi_{k'}^A$, and the scalar fields $\phi_{k'}$ are the respective tensor, vector, and scalar fields of the Lorentz algebra $so(d,1)$. The tensor fields  $\phi_{k'}^{AB}$ are symmetric and traceful. Now we are going to demonstrate that the field $\phi_0^{AB}$ enters description of spin-2 massless field, the fields $\phi_1^{AB}$, $\phi_1^A$ enter gauge invariant Stueckelberg description of spin-2 partial-massless field, while the fields $\phi_{k'}^{AB}$, $\phi_{k'}^A$, $\phi_{k'}$, $k'=2,\ldots,k$ enter gauge invariant Stueckelberg description of spin-2 massive fields. To this end we consider Lagrangian and gauge transformations.

\noindent {\bf Gauge invariant Lagrangian}. Lagrangian we found can be presented as
\beq
\label{04042014-01} (-\epsilon)^k \LL & = &  \LL_0 - \LL_1 +  \sum_{k'=2}^k (-)^{k'}  \LL_{k'}\,,
\\
\label{04042014-02} && \LL_0  =   \LL_0^2\,,
\\
\label{04042014-03} && \LL_1 = \LL_1^2 + \epsilon \LL_1^1\,,
\\
\label{04042014-04} && \LL_{k'} =  \LL_{k'}^2 + \epsilon \LL_{k'}^1 + \LL_{k'}^0\,,
\eeq
where we use the notation
\beq
\label{13042014-11} \frac{1}{e}\LL_{k'}^{2} & \equiv & \frac{1}{4}  \phi_{k'}^{AB}
(\DD^2 -m_{k'}^2 - 2\rho)\phi_{k'}^{AB} - \frac{1}{8} \phi_{k'}^{AA} (\DD^2 - m_{k'}^2 + 2\rho(d-2)) \phi_{k'}^{BB}  +   \half L_{k'}^A L_{k'}^A\,,\qquad
\\
\label{13042014-12} \frac{1}{e}\LL_{k'}^{1} & \equiv & \half  \phi_{k'}^A
(\DD^2 -m_{k'}^2 + d\rho)\phi_{k'}^A +  \half L_{k'} L_{k'}\,,
\\
\frac{1}{e}\LL_{k'}^{0} & \equiv & \half  \phi_{k'}
(\DD^2 -m_{k'}^2 + 2d\rho)\phi_{k'} \,,
\\
&& L_{k'}^A = \DD^B \phi_{k'}^{AB} - \half \DD^A \phi_{k'}^{BB} + |m_{k'}|\phi_{k'}^{A}\,,
\\
&& L_{k'} = \DD^B \phi_{k'}^B + \frac{\epsilon}{2}|m_{k'}| \phi_{k'}^{BB} + f_{k'} \phi_{k'}\,,
\\
&& f_{k'} \equiv \Bigl(\frac{2d}{d-1} |m_{k'}|^2 - 2 d |\rho| \Bigr)^{1/2}
\nonumber\\
&& \hspace{0.6cm}  = \,  \Bigl(\frac{2d(k'-1)(d-1-k')}{d-1}|\rho| \Bigr)^{1/2}\,,
\\
\label{04042014-05} && m_{k'}^2 = \rho k'(d-k')\,, \qquad k'=0,1,\ldots, k\,, \qquad k \equiv \frac{d-1}{2}\,.
\eeq

\noindent {\bf Gauge transformations}. To discuss gauge symmetries of the Lagrangian given in \rf{04042014-01}, we introduce the following gauge transformation
parameters:
\be \label{12042014-05}
\begin{array}{ccccccccccc}
\xi_0^A & & \xi_1^A & &\xi_2^A  & \ldots & \ldots && \xi_{k-1}^A & &
\xi_k^A
\\[12pt]
& \xi_1 & & \xi_2 & & \ldots & \ldots &\xi_{k-1}
& & \xi_k &
\end{array}
\ee
The gauge transformation parameters $\xi_{k'}^A$, $\xi_{k'}$ in \rf{12042014-05} are the respective vector and scalar fields of the Lorentz algebra $so(d,1)$. We note the following gauge transformations:
\beq
\label{04042014-06} && \delta \phi_{k'}^{AB} = \DD^A \xi_{k'}^B + \DD^B \xi_{k'}^A + \frac{2|m_{k'}|}{d-1} \eta^{AB} \xi_{k'}\,,
\\
\label{04042014-07} && \delta \phi_{k'}^A = \DD^A \xi_{k'} - \epsilon |m_{k'}| \xi_{k'}^A\,,
\\
\label{04042014-08} && \delta \phi_{k'} = - \epsilon f_{k'} \xi_{k'}\,.
\eeq

The following remarks are in order.

\noindent {\bf i}) Lagrangian $\LL_0$ in \rf{04042014-02} is invariant under $\xi_0^A$ gauge transformations given in \rf{04042014-06} when $k'=0$. This implies that the Lagrangian $\LL_0$ describes spin-2 massless field.

\noindent {\bf ii}) Lagrangian $\LL_1$ in \rf{04042014-03} is invariant under $\xi_1^A$ and $\xi_1$ gauge transformations given in \rf{04042014-06}, \rf{04042014-07} when $k'=1$. This implies that the Lagrangian $\LL_1$ describes spin-2 partial-massless field having square of mass parameter $m_1^2$ given in \rf{04042014-05}.

\noindent {\bf iii}) Lagrangian $\LL_{k'}$ in \rf{04042014-04} is invariant under $\xi_{k'}^A$ and $\xi_{k'}$ gauge transformations given in \rf{04042014-06}-\rf{04042014-08} when $k'=2,\ldots, k$. This implies that the Lagrangian $\LL_{k'}$ describes spin-2 massive field having square of mass parameter $m_{k'}^2$ given in \rf{04042014-05}.

\noindent {\bf iv}) From \rf{04042014-06}-\rf{04042014-08}, we see that all scalar and vector fields as well as trace of the tensor field $\phi_1^{AB}$ transform as Stueckelberg fields. In other words, just mentioned fields are realized as Stueckelberg fields in our description of spin-2 conformal field.

\noindent {\bf Summary}. Lagrangian of spin-2 conformal field in $(A)dS_{d+1}$  \rf{04042014-01} is a sum of Lagrangian $\LL_0$ \rf{04042014-02}, which  describes dynamics of spin-2 massless field, Lagrangian $\LL_1$ \rf{04042014-03}, which  describes dynamics of spin-2 partial-massless field with square of mass $m_1^2$ in \rf{04042014-05}, and Lagrangians $\LL_{k'}$, $k'=2,\ldots,k$, which describe dynamics of spin-2 massive fields with square of masses $m_{k'}^2$ \rf{04042014-05}.%
\footnote{ For arbitrary $d\geq 7$, little is known about interacting conformal gravities.  Study of local Weyl invariants in eight-dimensions may be found in Ref.\cite{Boulanger:2004zf}. Discussion of conformal supergravity in ten-dimensions may be found in Ref.\cite{Bergshoeff:1982az}.}

\noindent {\bf de Donder-like gauge}. Representation for Lagrangians in \rf{13042014-11},\rf{13042014-12} motivates us to introduce gauge condition which we refer to as de Donder-like gauge for spin-2 conformal field,
\beq
&& L_{k'}^A = 0 \,, \hspace{2cm}  k'=0,1,\ldots, k;
\nonumber\\[-10pt]
&& \hspace{8cm} \hbox{ de Donder-like gauge}\,.
\\[-10pt]
&& L_{k'}=0 \,, \hspace{2cm} k'=1,\ldots, k \,,
\nonumber
\eeq

%%%%%%%%%%%%%%%%%%%%%%%%%%%%%%%%%%%%%%%%%%%%%%%%%%%%%%%%%%%%%%%%%%%%%%%%%%%%%%%
%%%%%%%%%%%%%%%%%%%%%%%%%%%%%%%%%%%%%%%%%%%%%%%%%%%%%%%%%%%%%%%%%%%%%%%%%%%%%%%
\newsection{ \large Arbitrary spin-$s$ conformal field in $(A)dS_{d+1}$}\label{lagraarbspin}
%%%%%%%%%%%%%%%%%%%%%%%%%%%%%%%%%%%%%%%%%%%%%%%%%%%%%%%%%%%%%%%%%%%%%%%%%%%%%%%
%%%%%%%%%%%%%%%%%%%%%%%%%%%%%%%%%%%%%%%%%%%%%%%%%%%%%%%%%%%%%%%%%%%%%%%%%%%%%%%

{\bf Field content}. To develop ordinary-derivative and gauge invariant formulation of
spin-$s$ conformal field in $(A)dS_{d+1}$, for arbitrary odd $d\geq 3$, we use the following scalar,
vector, and tensor fields of the Lorentz algebra $so(d,1)$:
\beq
\label{04042014-09} && \phi_{k'}^{A_1\ldots A_{s'}} \,,\hspace{1.7cm}
k'=0,1,\ldots,k_s\,, \hspace{1.5cm} s'=\max(0,s-k'), \ldots, s;\qquad
\\
\label{04042014-10} && \hspace{3.2cm} k_s \equiv s + \frac{d-5}{2}\,.
\eeq
Tensor fields $ \phi_{k'}^{A_1\ldots A_{s'}}$ are totally symmetric and, when $s'\geq 4$, are double-traceless%
\footnote{ Discussion  of higher-spin field dynamics in terms of unconstrained fields can be found in  Refs.\cite{Sagnotti:2003qa,Buchbinder:2008ss}. Study of mixed-symmetry conformal field my be found in Ref.\cite{Vasiliev:2009ck} (see also Ref.\cite{Shaynkman:2004vu}). For interesting discussions of various aspects of mixed-symmetry fields see Refs.\cite{Alkalaev:2005kw}.}
\be \label{04042014-11}
\phi_{k'}^{AABB A_5\ldots A_{s'}} = 0\,,\qquad s'\geq 4\,.
\ee
The following remarks are in order.

\noindent {\bf i}) For $(A)dS_4$, fields in \rf{04042014-09} can be divided into two groups

\beq
 \label{04042014-12} && \phi_0^{A_1 \ldots A_s} \hspace{10cm} \hbox{massless}
\\
\label{04042014-14} && \phi_{k'}^{A_1\ldots A_{s'}} \,,\hspace{1cm}
k'=1,\ldots,s-1\,, \hspace{1cm} s'=s-k',\ldots, s;\hspace{1.2cm} \hbox{partial-massless}\qquad
\eeq
Below, we demonstrate that $\phi_0^{A_1 \ldots A_s}$ \rf{04042014-12} enters spin-$s$ massless field, while fields $\phi_{k'}^{A_1\ldots A_{s'}}$ in \rf{04042014-14} with $k'$ - fixed and $s'=s-k',\ldots,s$ enter gauge invariant Stueckelberg description of spin-$s$ partial-massless field having square of mass parameter $m_{k'}^2 = \rho k'(2s - 1 - k')$. To illustrate the field content given in \rf{04042014-12},\rf{04042014-14}, we use the shortcut $\phi_{k'}^{s'}$ for the field
$\phi_{k'}^{A_1\ldots A_{s'}}$ and note that, for $(A)dS_4$ and
arbitrary $s$, fields in \rf{04042014-12},\rf{04042014-14} can be presented as
in \rf{04042014-14x}.

\vspace{-0.3cm}
{\small
$$ \hbox{Field content for spin-$s$ conformal field in  $(A)dS_4$, $s$ - arbitrary } $$
\be \label{04042014-14x}
\begin{array}{ccccccccc}
\phi_0^s & & \phi_1^s & & \ldots & & \phi_{s-2}^s & &
\phi_{s-1}^s
\\[12pt]
& \phi_1^{s-1} & & \phi_2^{s-1} & \ldots & \phi_{s-2}^{s-1}
& & \phi_{s-1}^{s-1} &
\\[12pt]
& & \ldots  &   & \ldots & & \ldots  &  &
\\[12pt]
&  & & \phi_{s-2}^2  & & \phi_{s-1}^2 & & &
\\[12pt]
&  & &  & \phi_{s-1}^1& & & &
\end{array}
\ee }

\noindent {\bf ii}) For $(A)dS_{d+1}$, $d\geq 5$, fields in \rf{04042014-09} can be divided into the following three groups:
\beq
 \label{04042014-15} && \hspace{-0.8cm} \phi_0^{A_1 \ldots A_s} \hspace{10cm} \hbox{massless}
\\
\label{04042014-16} && \hspace{-0.8cm} \phi_{k'}^{A_1\ldots A_{s'}} \,,\hspace{1cm}
k'=1,\ldots,s-1\,, \hspace{1cm} s'=s-k',\ldots, s;\hspace{1.2cm} \hbox{partial-massless}\qquad
\\
\label{04042014-17} && \hspace{-0.8cm} \phi_{k'}^{A_1\ldots A_{s'}} \,,\hspace{1cm}
k'=s,\ldots,k_s\,, \hspace{1.5cm} s'=0,1,\ldots, s;\hspace{1.7cm} \hbox{massive}\qquad
\eeq
Below we show that {\bf a}) the field $\phi_0^{A_1 \ldots A_s}$ \rf{04042014-15} enters spin-$s$ massless field; {\bf b}) the fields $\phi_{k'}^{A_1\ldots A_{s'}}$ \rf{04042014-16} with $k'$ - fixed and $s'=s-k',\ldots,s$ enter gauge invariant Stueckelberg description of spin-$s$ partial-massless field with square of mass parameter $m_{k'}^2 = \rho k'(2s + d- 4 - k')$; {\bf c}) the fields $\phi_{k'}^{A_1\ldots A_{s'}}$ \rf{04042014-17} with $k'$ - fixed and $s'=0,1,\ldots,s$ enter gauge invariant Stueckelberg description of spin-$s$ massive field having square of mass parameter $m_{k'}^2 = \rho k'(2s + d- 4 - k')$. To illustrate the field content in \rf{04042014-15}-\rf{04042014-17}, we use the shortcut $\phi_{k'}^{s'}$ for the field $\phi_{k'}^{A_1\ldots A_{s'}}$ and note that, for $d\geq 5$ and arbitrary $s$, fields in \rf{04042014-15}-\rf{04042014-17} can be presented as in \rf{04042014-18}.

\vspace{-0.6cm}
{\small
\beq
\label{04042014-18} && \hspace{-1cm} \hbox{Field content of spin-$s$ conformal field in  $(A)dS_{d+1}$ for odd $d \geq 5 $, \ $s$ - arbitrary, $k_s\equiv s + \frac{d-5}{2}$ }
\nonumber\\
&&\hspace{-1cm} \phi_0^s \hspace{1cm} \phi_1^s
\hspace{1cm} \ldots \hspace{1cm}\ldots \hspace{1cm} \ldots
\hspace{1cm} \ldots \hspace{1cm}\ldots \hspace{1.8cm} \phi_{k_s-1}^s
\hspace{1cm}  \phi_{k_s}^s \qquad
\nonumber\\[15pt]
&& \hspace{-0.3cm} \phi_1^{s-1} \hspace{1cm}
\phi_2^{s-1} \hspace{1cm} \ldots \hspace{1cm} \ldots
\hspace{1cm}\ldots \hspace{1cm}\ldots \hspace{1cm}
\phi_{k_s-1}^{s-1} \hspace{1cm} \phi_{k_s}^{s-1}
\nonumber\\[14pt]
&& \hspace{0.5cm} \ldots \hspace{1cm}  \ldots \hspace{1cm}  \ldots
\hspace{1cm} \ldots \hspace{1cm} \ldots   \hspace{1cm} \ldots
\hspace{1cm} \ldots \hspace{1cm} \ldots
\\[14pt]
&& \hspace{1cm}  \phi_{s-1}^1 \hspace{1cm} \phi_s^1
\hspace{1cm}  \ldots  \hspace{1cm} \ldots \hspace{1.5cm}
\phi_{k_s-1}^1 \hspace{0.8cm} \phi_{k_s}^1
\nonumber\\[15pt]
&& \hspace{1.8cm} \phi_s^0 \hspace{1.2cm} \phi_{s+1}^0
\hspace{1.2cm} \ldots \hspace{1cm} \phi_{k_s-1}^0 \hspace{1cm}
\phi_{k_s}^0
\nonumber
\eeq }

\noindent {\bf iii}) The lowest value of $d$ when the
scalar fields appear in the field content is given by $d=5$. Namely, for $(A)dS_6$ and
arbitrary $s$, the field content in \rf{04042014-18} is simplified as
{\small
$$ \hbox{Field content for spin-$s$ conformal field in  $(A)dS_6$,  $s$ - arbitrary   }$$
\be
\begin{array}{ccccccccc}
\phi_0^s & & \phi_1^s & & \ldots & & \phi_{s-1}^s & &
\phi_s^s
\\[12pt]
& \phi_1^{s-1} & & \phi_2^{s-1} & \ldots & \phi_{s-1}^{s-1}
& & \phi_s^{s-1} &
\\[12pt]
& & \ldots  &   & \ldots & & \ldots  &  &
\\[12pt]
&  & & \phi_{s-1}^1  & & \phi_s^1 & & &
\\[12pt]
&  & &  & \phi_s^0& & & &
\end{array}
\ee }

\noindent {\bf Generating form of field content}. To streamline the presentation of our gauge invariant ordinary-derivative formulation we use the oscillators  $\alpha^A$, $\zeta$,
$\vartheta$, $\chi$, and collect scalar, vector and tensor fields
\rf{04042014-09} into the ket-vector $\phik$ defined by%
\footnote{ In earlier literature, extensive use of oscillator formalism may
be found in Ref.\cite{Vasiliev:1987tk} (see also Ref.\cite{Bekaert:2006ix}).}
\beq
\label{04042014-19} && \phik = \sum_{k'=0}^{k_s} \frac{\vartheta^{k_s-k'}}{\sqrt{(k_s-k')!}} |\phi_{k'}\rangle \,,
\\
\label{04042014-20} && |\phi_{k'}\rangle = \sum_{s'=\max(0,s-k')}^s \frac{ \chi^{k'+s'-s} \zeta^{s-s'} \alpha^{A_1}\ldots \alpha^{A_{s'}} }{s'!\sqrt{(k'+s'-s)!(s-s')!}}\phi_{k'}^{A_1\ldots A_{s'}}\,.
\eeq
From \rf{04042014-11},\rf{04042014-19},\rf{04042014-20} , we see that the ket-vectors
$\phik$, $|\phi_{k'}\rangle$ satisfy the constraints
\beq
\label{04042014-21} && (N_\alphabf + N_\zeta)\phik =  s\phik\,,
\\
\label{04042014-22} && (N_\zeta + N_\vartheta + N_\chi)\phik = k_s \phik\,,
\\
\label{04042014-23} && (N_\zeta +  N_\chi) |\phi_{k'} \rangle  = k' |\phi_{k'}\rangle \,,
\\
\label{04042014-23x} && (\bar\alphabf^2)^2 \phik = 0\,.
\eeq
From \rf{04042014-21}, we learn that the ket-vector $\phik$ is
degree-$s$ homogeneous polynomial in the oscillators $\alpha^A$, $\zeta$, while constraint \rf{04042014-22} tells us that $\phik$ degree-$k_s$ homogeneous polynomial in the oscillators $\zeta$, $\vartheta$, $\chi$. Constraint \rf{04042014-23}
implies that the ket-vectors $|\phi_{k'}\rangle$ is the
degree-$k'$ homogeneous polynomials in the oscillators
$\zeta$, $\chi$. Constraint \rf{04042014-23x} is just the presentation of the double-tracelessness constraints \rf{04042014-11} in terms of the ket-vector $\phik$.

We now proceed with the discussion of gauge invariant Lagrangian in the framework of our ordinary-derivative approach. We would like to discuss generating form and component form of the Lagrangian. We discuss these two representations for the Lagrangian in turn.

\noindent {\bf Gauge invariant Lagrangian. Generating form}. The Lagrangian we found is given by
\beq
\label{04042014-24} \LL  &  = &  \half  e \phibr  E \phik \,,
\\
\label{04042014-25} E & = & \bigl(1-\frac{1}{4}\alphabf^2 \bar\alphabf^2\bigr) \bigl(\Box_{_\pAdS} + m_1 + \rho
\alphabf^2\bar\alphabf^2\bigr) - \Lbf \bar\Lbf\,,
\\
&& m_1 = -m^2 +\rho \Bigl( s(s+d-5) -2d+ 4 +N_\zeta(2s+d-1-N_\zeta)\Bigr)\,,
\\
\label{04042014-28} && m^2 = \rho N_{\zeta \chi}(2s+d-4- N_{\zeta \chi})\,, \qquad N_{\zeta \chi} \equiv N_\zeta + N_\chi\,,
\\
\label{04042014-26} && \bar\Lbf \equiv  \bar\alphabf \Dbf - \half \alphabf \Dbf  \bar\alphabf^2  -
\eb_1\Pibf^\smponetwo + \half e_1 \bar\alphabf^2\,,
\\
\label{04042014-27} && \Lbf \equiv \alphabf \Dbf  - \half \alphabf^2 \bar\alphabf \Dbf  - e_1 \Pibf^\smponetwo + \half \eb_1 \alphabf^2\,,
\\
\label{04042014-27x} && e_1 =   \zeta \ewt_1 \chib \,, \qquad \eb_1 = - \chi  \ewt_1\bar\zeta\,,
\\
&& \ewt_1 \equiv \bigl(|\rho|(2s+d-5-2N_\zeta - N_\chi)\bigr)^{1/2} e_\zeta   \,,
\\
\label{04042014-27xy} && e_\zeta \equiv \Bigl(\frac{2s+d-3-N_\zeta}{2s+d-3-2N_\zeta} \Bigr)^{1/2} \,.
\eeq
Definition of operators appearing in \rf{04042014-25}-\rf{04042014-27xy} may be found in Appendix. We note that two-derivative part of the operator $E$ \rf{04042014-25} coincides with the standard Fronsdal operator represented in terms of the oscillators. Operator $E$ \rf{04042014-25} can also be represented as
\beq
\label{04042014-29} E & = & \Box_\pAdS + M_1 -  \frac{1}{4}\alphabf^2 \bar\alphabf^2 ( \Box_\pAdS + M_2)  - \Lbf \bar\Lbf\,,
\\
&& M_1 = -m^2 +\rho \Bigl( s(s+d-5) -2d+ 4 +N_\zeta(2s+d-1-N_\zeta)\Bigr)\,,
\\
&& M_2 = -m^2 +\rho \Bigl( s(s+d-1) - 6 + N_\zeta(2s+d-5-N_\zeta)\Bigr)\,.
\eeq

\noindent {\bf Gauge symmetries}. To discuss gauge symmetries of the Lagrangian in \rf{04042014-24} we use the gauge transformation parameters given by
\be \label{04042014-30}
\xi_{k'}^{A_1\ldots A_{s'}} \,,\hspace{1.7cm}
k'=0,1,\ldots,k_s\,, \hspace{1.5cm} s'=\max(0,s-1-k'), \ldots, s-1;\qquad
\ee
where $k_s$ is defined in \rf{04042014-10}. The gauge transformation parameters in \rf{04042014-30} are scalar, vector, and tensor fields of the Lorentz algebra $so(d,1)$. The gauge transformation parameters $ \xi_{k'}^{A_1\ldots A_{s'}}$ are totally symmetric and, when $s'\geq 2$, are traceless
\be \label{04042014-31}
\xi_{k'}^{BB A_3\ldots A_{s'}} = 0\,, \qquad s'\geq 2\,.
\ee
As usually, to simplify the presentation, we collect the gauge transformation parameters into a ket-vector defined by
\beq
\label{04042014-32} && \xik = \sum_{k'=0}^{k_s} \frac{\vartheta^{k_s-k'}}{\sqrt{(k_s-k')!}} |\xi_{k'}\rangle \,,
\\
\label{04042014-33} && |\xi_{k'}\rangle = \sum_{s'=\max(0,s-1-k')}^{s-1} \frac{ \chi^{k'+s'-s+1} \zeta^{s-1-s'} \alpha^{A_1}\ldots \alpha^{A_{s'}} }{s'!\sqrt{(k'+s'-s+1)!(s-s'-1)!}}\xi_{k'}^{A_1\ldots A_{s'}}\,.
\eeq
From \rf{04042014-31}-\rf{04042014-33} , we see that the ket-vectors
$\xik$, $|\xi_{k'}\rangle$ satisfy the constraints
\beq
\label{04042014-34} && (N_\alphabf + N_\zeta)\xik =  (s-1)\xik\,,
\\
\label{04042014-35} && (N_\zeta + N_\vartheta + N_\chi)\xik = k_s \xik\,,
\\
\label{04042014-36} && (N_\zeta +  N_\chi) |\xi_{k'} \rangle  = k' |\xi_{k'}\rangle \,,
\\
\label{04042014-37}  && \bar\alphabf^2 \xik = 0\,.
\eeq
From \rf{04042014-34}, we learn that the ket-vector $\xik$ is a
degree-$(s-1)$ homogeneous polynomial in the oscillators $\alpha^A$, $\zeta$,  while  constraint \rf{04042014-35} tells us that the $\xik$ is a degree-$k_s$ homogeneous polynomial in the oscillators $\zeta$, $\vartheta$, $\chi$. Constraints \rf{04042014-36}
implies that the ket-vector $|\xi_{k'}\rangle$ is a
degree-$k'$ homogeneous polynomial in the oscillators
$\zeta$, $\chi$. Constraint \rf{04042014-37} is just the presentation of the tracelessness  constraints \rf{04042014-31} in terms of the ket-vector $\xik$.

Using ket-vectors $\phik$, $\xik$, gauge transformation for spin-$s$ conformal field can be presented as
\be \label{12042014-04}
\delta \phik   =   G \xik\,, \qquad  G \equiv \alphabf \Dbf - e_1 - \alphabf^2 \frac{1}{2N_\alphabf +d-1}\eb_1\,,
\ee
where operators $e_1$, $\eb_1$ are  given in \rf{04042014-27x}.

\noindent {\bf Component form of Lagrangian and gauge transformations}. For deriving the component form of Lagrangian it is convenient to use representation for the Lagrangian with the operator $E$ given in \rf{04042014-29}. By plugging ket-vector
\rf{04042014-19} into \rf{04042014-24} we obtain the component
form of the  Lagrangian,
\beq
\label{05042014-01} && \LL = (-\epsilon)^{k_s}\sum_{k'=0}^{k_s} (-)^{k'} \LL_{k'}\,,
\\
\label{05042014-02} && \LL_{k'} \equiv \sum_{s'= \max{(0,s-k')}}^s \epsilon^{s'-s} \LL_{k'}^{s'}\,,
\eeq
where we use the notation
\beq
\label{05042014-03} \frac{1}{e} \LL_{k'}^{s'} & \equiv & \frac{1}{2 s'!} \phi_{k'}^{A_1\ldots A_{s'}}
(\DD^2 + M_{1,k'}^{s'})\phi_{k'}^{A_1\ldots A_{s'}}
- \frac{1}{8(s'-2)!} \phi_{k'}^{BBA_3\ldots A_{s'}} (\DD^2 + M_{2,k'}^{s'})
\phi_{k'}^{EEA_3\ldots A_{s'}}\qquad
\nonumber\\
& + &  \frac{1}{2 (s'-1)!} L_{k'}^{A_1\ldots A_{s'-1}} L_{k'}^{A_1\ldots
A_{s'-1}}\,,
\\
\label{05042014-04} && L_{k'}^{A_1\ldots A_{s'-1}} \equiv
\DD^B \phi_{k'}^{A_1\ldots A_{s'-1} B}
- \frac{s'-1}{2}\DD^{(A_1} \phi_{k'}^{A_2\ldots A_{s'-1}) BB}
\nonumber\\
&& \hspace{1.8cm} + \, f_{k'}^{s'}  \phi_{k'}^{\tr\ A_1\ldots A_{s'-1}} +
\frac{\epsilon}{2} f_{k'}^{s'+1}   \phi_{k'}^{A_1\ldots A_{s'-1}BB}\,,
\\
&& \phi_{k'}^{\tr\ A_1\ldots A_{s'}} \equiv \phi_{k'}^{A_1\ldots A_{s'}} -
\frac{s'(s'-1)}{2(2s'+d-3)} \eta^{(A_1A_2} \phi_{k'}^{A_3\ldots A_{s'})
BB}\,,
\eeq

\beq
\label{13042014-28} && M_{1,k'}^{s'} \equiv -m_{k'}^2 + \rho \bigl(2(s-1)(s+d-2) - s'(s'+d-1)\bigr)\,,
\\
&& M_{2,k'}^{s'} \equiv  -m_{k'}^2 + \rho \bigl( 2s(s+d-3) - 6  - s'(s'+d-5) \bigr)\,,
\\
\label{05042014-05} && m_{k'}^2 = \rho k'(2s+d-4 -k')\,,
\\
\label{05042014-06} && f_{k'}^{s'}  \equiv \Bigl(\epsilon \frac{(s-s'+1) (s+s'+d-3)}{2s'+d-3} \bigl( m_{k'}^2 - \rho (s-s')(s+s'+d-4) \bigr)\Bigr)^{1/2}\,.
\eeq
We recall that $e=\det e_\mu^A$, $e_\mu^A$ stands for vielbein of (A)dS, $\DD^2 \equiv \DD^A \DD^A$, where $\DD^A$ stands for the covariant derivative in (A)dS.  We use $\rho= \epsilon/R^2$, where $\epsilon = 1 (-1)$ for dS (AdS) and $R$ is radius of (A)dS. Quantities $L_{k'}^{A_1\ldots A_{s'-1}}$ \rf{05042014-04} are referred to as de Donder divergences in this paper. Note that using $m_{k'}^2$ given in \rf{05042014-05} allows us to represent $f_{k'}^{s'}$ in the following factorized form:
\be
f_{k'}^{s'}  \equiv \Bigl(|\rho| \frac{(s-s'+1) (s+s'+d-3) (k'+s'-s) (s+s'+d-4-k')}{2s'+d-3}\Bigr)^{1/2}\,.
\ee

\noindent {\bf Component form of gauge transformations}. Component form of gauge transformations is easily found by plugging ket-vectors $\phik$, $\xik$ into \rf{12042014-04}. Doing so, we get
\be \label{12042014-06}
\delta \phi_{k'}^{A_1\ldots A_{s'}} = s' \DD^{(A_1}\xi_{k'}^{A_2\ldots A_{s'})} - \epsilon f_{k'}^{s'+1} \xi_{k'}^{A_1\ldots A_{s'}} + \frac{s'(s'-1)f_{k'}^{s'}}{2s'+d-5} \eta^{(A_1A_2} \xi_{k'}^{A_3 \ldots A_{s'})}\,.
\ee

The following remarks are in order.

\noindent {\bf i}) Lagrangian $\LL_0$ in \rf{05042014-02} is invariant under $\xi_0^{A_1\ldots A_{s-1}}$ gauge transformations given in \rf{12042014-06} when $k'=0$. This implies that the Lagrangian $\LL_0$ describes spin-$s$ massless field.

\noindent {\bf ii}) For $k'=1,\ldots,s-1$, Lagrangian $\LL_{k'}$ \rf{05042014-02} is invariant under $\xi_{k'}^{A_1\ldots A_{s'}}$, $s'= s-k'-1, \ldots, s-1$, gauge transformations given in \rf{12042014-06}. This implies that the Lagrangian $\LL_{k'}$ describes spin-$s$ partial-massless field having square of mass parameter $m_{k'}^2$ given in \rf{05042014-05}.

\noindent {\bf iii}) For $k'=s,\ldots,k_s$, Lagrangian $\LL_{k'}$ \rf{05042014-02} is invariant under $\xi_{k'}^{A_1\ldots A_{s'}}$, $s'= 0, \ldots, s-1$, gauge transformations given in \rf{12042014-06}. This implies that the Lagrangian $\LL_{k'}$ describes spin-$s$ massive field having square of mass parameter $m_{k'}^2$ given in \rf{05042014-05}.

\noindent {\bf iv}) Using \rf{12042014-06}, one can make sure, that all scalar fields, all vector fields, traceless parts  of tensor fields $\phi_{k'}^{A_1\ldots A_{s'}}$, $s'=2,\ldots, s-1$, $k'=s-s',\ldots,k_s$, and traces of tensor fields $\phi_{s+1-s'}^{A_1\ldots A_{s'}}$, $s'=2,\ldots, s$, transform as a Stueckelberg fields. In other words, just mentioned fields are realized as Stueckelberg fields in our description of spin-$s$ conformal field.

\noindent {\bf v}) Note that we express our gauge invariant Lagrangian in terms of de Donder divergences given in \rf{05042014-04}. Obviously it is the use of de Donder divergences that allows us to simplify considerably the gauge invariant Lagrangian.%
\footnote{ Representation of gauge invariant Lagrangian for massless, partial-massless, and  massive (A)dS fields in terms of de Donder-like divergences was found in Ref.\cite{Metsaev:2008ks,Metsaev:2009hp}. Alternative representation of gauge invariant Lagrangian for partial-massless and  massive (A)dS fields was obtained for the first time in Ref.\cite{Zinoviev:2001dt}. Study of partial-massless fields in frame-like approach may be found in Ref.\cite{Skvortsov:2006at}. In the framework of tractor and BRST approaches, partial-massless fields were studied in Refs.\cite{Alkalaev:2011zv}. Interacting partial-massless fields are studied in Refs.\cite{Joung:2012hz} (see also Refs.\cite{Deser:2012qg,Deser:2013bs}.}

\noindent {\bf Summary}. Lagrangian of spin-$s$ conformal field in $(A)dS_{d+1}$ given in \rf{05042014-01} is a sum of Lagrangian $\LL_0$ \rf{05042014-02}, which  describes dynamics of spin-$s$ massless field, Lagrangians $\LL_{k'}$ \rf{05042014-02}, $k'=1,\ldots, s-1$, which  describe dynamics of spin-$s$ partial-massless fields, and Lagrangians $\LL_{k'}$, \rf{05042014-02}, $k'=s,\ldots, k_s$, which describe dynamics of spin-$s$ massive fields.
Square of mass parameter for massless, partial-massless, and massive fields is described on equal footing by $m_{k'}^2$ given in \rf{05042014-05}. Our result for $m_{k'}^2$ confirms the conjecture about $m_{k'}^2$ made in Ref.\cite{Tseytlin:2013jya}.

\noindent {\bf de Donder-like gauge}. Representation for Lagrangians in \rf{04042014-24},\rf{04042014-25} motivates us to introduce gauge condition which we refer to as de Donder-like gauge for spin-$s$ conformal field,
\be \label{13042014-14}
\bar\Lbf \phik = 0\,,  \hspace{1cm} \hbox{ de Donder-like gauge},
\ee
where the operator $\bar\Lbf$ is given in \rf{04042014-26}. It is easy to see that gauge \rf{13042014-14} considerably simplifies the Lagrangian in \rf{04042014-24}.%
\footnote{ For massless fields, discussion of the standard de Donder-Feynman
gauge may be found in
Refs.\cite{Guttenberg:2008qe}.
Extensive use of Donder-like gauge conditions for studying the $AdS/CFT$ correspondence may be found in  Refs.\cite{Metsaev:2009ym}-\cite{Metsaev:2010zu}. We  believe that our de Donder-like gauge might also be useful for the study of AdS/FT correspondence along the lines in Refs.\cite{Tseytlin:2013jya,Tseytlin:2013fca}. Also we note that our approach to conformal (A)dS fields streamlines application of general methods for the computation of one-loop effective action developed in Refs.\cite{Fradkin:1981iu,Barvinsky:1985an}.
}
In terms of tensor fields, gauge \rf{13042014-14} can be presented as
\be \label{13042014-29}
L_{k'}^{A_1\ldots A_{s'}} = 0  \,,\hspace{1cm}
k'=0,1,\ldots,k_s\,, \hspace{1cm} s'=\max(0,s-1-k'), \ldots, s-1;\qquad
\ee
where $L_{k'}^{A_1\ldots A_{s'}}$ is given in \rf{05042014-04}.

\noindent {\bf Left-over gauge symmetries of de Donder-like gauge}. We note that de Donder-like gauge has left-over gauge symmetry. This symmetry can easily be obtained by using the following relation
\beq
\label{13042014-25} && \bar\Lbf G\xik =   \bigl( \Box_{_\pAdS} + M_{_\FP} \bigr) \xik\,,
\\
&& M_{_\FP} \equiv - m^2  + \rho \Bigl( (s-1)(s+d-2) + N_\zeta(2 s + d - 3 -N_\zeta)  \Bigr) \,,
\eeq
where $m^2$ is given in \rf{04042014-28}, while $G$ stands for operator entering gauge transformation in \rf{12042014-04}. From \rf{13042014-25}, we see gauge condition \rf{13042014-14} is invariant under left-over gauge transformation provided the gauge transformation parameter satisfies the equation
\be \label{13042014-26}
\bigl( \Box_{_\pAdS} +  M_{_\FP} \bigr)\xik = 0\,.
\ee
In terms of tensor fields, Eq.\rf{13042014-26} can be represented as
\be \label{13042014-27}
(\DD^2 + M_{1,k'}^{s'})\xi_{k'}^{A_1\ldots A_{s'}} = 0  \,,\hspace{0.6cm}
k'=0,1,\ldots,k_s\,, \hspace{0.6cm} s'=\max(0,s-1-k'), \ldots, s-1;
\ee
where $M_{1,k'}^{s'}$ is given in \rf{13042014-28}. We note that the kinetic operators  appearing in \rf{13042014-27} are the kinetic operators of Faddeev-Popov fields when we use de Donder-like gauge given in \rf{13042014-14}.

\noindent {\bf Partition function of conformal field via de Donder-like gauge}. We now explain how the partition function of conformal field obtained in Ref.\cite{Tseytlin:2013jya} arises in the framework of our approach. Using gauge-fixed Lagrangian, partition function of conformal field, denoted by $Z^\total$, can be presented as (for details of the derivation \rf{13042014-17}, see below):
\beq \label{13042014-17}
&& Z^\total = \prod_{k'=0}^{k_s} Z_{k'}\,,
\\
\label{13042014-18} && Z_{k'}  \equiv  \prod_{s'= \max(0,s-1-k')}^{s-1} D_{k'}^{s'} D_{k'}^{s'} \Bigl/ \prod_{s'= \max(0,s-k')}^s D_{k'}^{s'}\prod_{s'= \max(0,s-2-k')}^{s-2} D_{k'}^{s'}\,,
\\[5pt]
\label{13042014-19} && \hspace{2cm} D_{k'}^{s'} \equiv \bigl(\det(-\DD^2 - M_{1,k'}^{s'}) \bigr)^{1/2}\,,
\eeq
where, in \rf{13042014-19}, the determinant is evaluated on space of traceless rank-$s'$ tensor field. In \rf{13042014-17}, the $Z_{k'}$ is partition function of spin-$s$ (A)dS field having square of mass parameter $m_{k'}^2$.  It is easy to see that $Z_{k'}$ \rf{13042014-18} take the form
\beq
\label{13042014-20a1} && Z_{k'} = \frac{D_{k'}^{s-1}D_{k'}^{s-1-k'}}{D_{k'}^sD_{k'}^{s-2-k'}}\,, \hspace{2cm} \hbox{ for } \ \ k' = 0,1, \ldots, s-2;
\\
\label{13042014-20a2} && Z_{k'} = \frac{D_{k'}^{s-1} D_{k'}^0}{D_{k'}^s}\,, \hspace{2.8cm} \hbox{ for } \ \ k' = s-1;
\\
\label{13042014-21a3} && Z_{k'} = \frac{D_{k'}^{s-1}}{D_{k'}^s}\,, \hspace{3.4cm} \hbox{ for } \ \ k' = s,s+1, \ldots, k_s\,.
\eeq
We now note the well-known relation for $D_{k'}^{s'}$ in \rf{13042014-19}
\be \label{13042014-19x}
D_{k'}^{s'} = D_{k'}^{s'\perp} D_{k'}^{s'-1}\,,
\ee
where, in \rf{13042014-19x}, the $D_{k'}^{s'\perp}$ takes the same form as in \rf{13042014-19}, while the determinant is evaluated on space of divergence-free ($\DD^A \phi^{AA_2\ldots A_{s'}}= 0$) traceless rank-$s'$ tensor field. Using \rf{13042014-19x}, the $Z_{k'}$ in \rf{13042014-20a1}-\rf{13042014-21a3} can be represented as
\beq
\label{13042014-20} && Z_{k'} = \frac{D_{k'}^{s-1-k'\perp}}{D_{k'}^{s\perp}}\,, \hspace{2cm} \hbox{ for } \ \ k' = 0,1, \ldots, s-1;
\\
\label{13042014-21} && Z_{k'} = \frac{1}{D_{k'}^{s\perp}}\,, \hspace{2.8cm} \hbox{ for } \ \ k' = s,s+1, \ldots, k_s\,.
\eeq
Plugging \rf{13042014-20},\rf{13042014-21} into \rf{13042014-17}, we get
\be \label{13042014-22}
Z^\total = \prod_{k'=0}^{s-1} \frac{D_{k'}^{s-1-k'\perp}}{D_{k'}^{s\perp}} \prod_{k'=s}^{k_s} \frac{1}{D_{k'}^{s\perp}}\,.
\ee
Taking into account values of the mass parameters entering $D_{k'}^{s-1-k'}$, $D_{k'}^s$ \rf{13042014-19},
\beq
\label{13042014-23} && M_{1,k'}^{s-1-k'} =  \rho k' + \rho (s-1)(s+d-2)\,,
\\
\label{13042014-24} && M_{1,k'}^s = - \rho s + \rho (s-2 -k')(s+d-2-k')\,,
\eeq
we verify that our expression for $Z^\total$ in \rf{13042014-22} coincides with the one in Ref.\cite{Tseytlin:2013jya}.%
\footnote{ Note that our labels differ from the ones in  Ref.\cite{Tseytlin:2013jya}.}

We now present some details of the derivation of partition function \rf{13042014-17}. To this end we note that the de Donder-like gauge condition implies the use of the following gauge fixing term:
\be \label{07062014-01}
\frac{1}{e} \LL_{\rm g.fix} =  \half \phibr \Lbf \bar\Lbf \phik +  \langle \cb| ( \Box_{_\pAdS} + M_{_\FP} \bigr) \ck\,,
\ee
where $|\cb \rangle $ and $\ck$ stand for ket-vectors of the Faddeev-Popov fields. Decompositions of the ket-vectors $|\cb \rangle $, $\ck$  into the respective tensor fields $\cb_{k'}^{A_1\ldots A_{s'}}$, $c_{k'}^{A_1\ldots A_{s'}}$ take the same form as the decomposition of the ket-vector $\xik$ into the tensor fields $\xi_{k'}^{A_1\ldots A_{s'}}$ in \rf{04042014-32},\rf{04042014-33}.  Also, the ket-vectors $|\cb \rangle $, $\ck$  satisfy the same algebraic constraints as the ket-vector $\xik$ in   \rf{04042014-34}-\rf{04042014-37}. Use of \rf{04042014-24} and \rf{07062014-01}, leads to the following gauge-fixed Lagrangian:
\beq
\label{07062014-02} && \hspace{-0.5cm} \LL_\total = \LL + \LL_{\rm g.fix}\,,
\\
\label{07062014-03} && \hspace{-0.5cm} \frac{1}{e} \LL_\total = \half \phibr \bigl( 1 - \frac{1}{4}\alphabf^2\bar\alphabf^2 \bigr) \bigl( \Box_{_\pAdS} + m_1 + \rho \alphabf^2\bar\alphabf^2 \bigr)\phik  + \langle \cb| ( \Box_{_\pAdS} + M_{_\FP} \bigr) \ck\,.\qquad
\eeq
Gauge-fixed Lagrangian \rf{07062014-03} can alternatively be represented as
\be \label{07062014-04}
\frac{1}{e} \LL_\total = \half \phibr \Bigl(\Box_{_\pAdS} + M_1 -  \frac{1}{4}\alphabf^2 \bar\alphabf^2 ( \Box_{_\pAdS} + M_2)\Bigr) \phik  + \langle \cb| ( \Box_{_\pAdS} + M_{_\FP} \bigr) \ck\,.
\ee
Another representation of gauge-fixed Lagrangian \rf{07062014-03} is obtained by using the decomposition of double-traceless ket-vector $\phik$ \rf{04042014-19} into two traceless ket-vectors
defined by the relation
\beq
\label{29042014-01}   \phik & = &|\phi_{_\I} \rangle   + \alphabf^2 \NN |\phi_{_\II} \rangle\,, \qquad \bar\alphabf^2 |\phi_{_\I} \rangle =0\,,  \qquad \bar\alphabf^2 |\phi_{_\II} \rangle =0\,,
\\
\label{29042014-02} && \NN\equiv ((2s+d-3-2N_\zeta)(2s+d-5-2N_\zeta))^{-1/2}\,.
\eeq
Using \rf{29042014-01}, we represent gauge-fixed Lagrangian \rf{07062014-04} as
\be \label{07062014-05}
\frac{1}{e} \LL_\total =  \half \langle \phi_{_\I} | (\Box_{_{(A)dS}} + M_1) |\phi_{_\I} \rangle - \half \langle \phi_{_\II} | (\Box_{_{(A)dS}} + M_2 + 4\rho ) |\phi_{_\II} \rangle
+  \langle \cb |\bigl( \Box_{_{(A)dS}} + M_{_\FP} \bigl) |c\rangle\,.
\ee
It is the representation given in \rf{07062014-05} that leads immediately to partition function \rf{13042014-17}-\rf{13042014-19}. To see this it is convenient to write down a component form of Lagrangian in \rf{07062014-05}. For that purpose we note that a decomposition of the ket-vector $|\phi_{_\I} \rangle$  into tensor fields $\phi_{\I\, k'}^{A_1\ldots A_{s'}}$ takes the same form as the decomposition of the ket-vector $\phik$ into tensor fields $\phi_{k'}^{A_1\ldots A_{s'}}$ in \rf{04042014-19},\rf{04042014-20}, while a decomposition of the ket-vector $|\phi_{_\II} \rangle$  into tensor fields $\phi_{\II\, k'}^{A_1\ldots A_{s'}}$ takes the form as the decomposition of ket-vector $\phik$ into tensor fields $\phi_{k'}^{A_1\ldots A_{s'}}$ in \rf{04042014-19},\rf{04042014-20} with  the replacement $s\rightarrow s-2$ in \rf{04042014-20}. This is to say that component form  of gauge-fixed Lagrangian \rf{07062014-05} takes the form
\beq
\label{07062014-06}  && \LL_\total = (-\epsilon)^{k_s}\sum_{k'=0}^{k_s} (-)^{k'} \LL_{\total\, k'}\,,
\\
&&   \LL_{\total\, k'}=\!\!  \sum_{s'= \max{(0,s-k')}}^s \hspace{-0.7cm} \epsilon^{s'-s} \LL_{\I\,k'}^{s'} - \!\! \sum_{s'= \max{(0,s-2-k')}}^{s-2} \hspace{-0.7cm} \epsilon^{s'-s} \LL_{\II\,k'}^{s'} + \!\! \sum_{s'= \max{(0,s-1-k')}}^{s-1} \hspace{-0.7cm} \epsilon^{s'+1-s} \LL_{\FP\,k'}^{s'}\,,\qquad
\\
\label{07062014-07}  &&     \LL_{\I\, k'}^{s'}  = \frac{e}{2s'!}\, \phi_{\I\,k'}^{A_1 \ldots A_{s'}} (\DD^2 + M_{1\,k'}^{s'})  \phi_{\I\,k'}^{A_1 \ldots A_{s'}}\,,
\\
\label{07062014-08} &&   \LL_{\II\,k'}^{s'}  =  \frac{e}{2s'!}\, \phi_{\II\,k'}^{A_1 \ldots A_{s'}} (\DD^2 + M_{1\,k'}^{s'})  \phi_{\II \,k'}^{A_1 \ldots A_{s'}}\,,
\\
\label{07062014-09} &&   \LL_{\FP\,k'}^{s'}  = \frac{e}{s'!}\, \cb_{k'}^{A_1 \ldots A_{s'}} (\DD^2 + M_{1\,k'}^{s'})  c_{k'}^{A_1 \ldots A_{s'}}\,.
\eeq
From \rf{07062014-07}-\rf{07062014-09}, we see that $\LL_\total$ \rf{07062014-06} leads to the partition function given in \rf{13042014-17}-\rf{13042014-19}.

As remark, we note that gauge-fixed Lagrangian \rf{07062014-03} is invariant under global BRST transformations given by
\be
\delta_{_\BRST} \phik = \lambda G \ck\,, \qquad
\delta_{_\BRST} |\cb\rangle  = \lambda  \bar\Lbf \phik \,,\qquad  \delta_{_\BRST} \ck  =  0  \,,
\ee
where $\lambda$ is odd parameter of the global BRST transformations. We assume the following hermitian conjugation rules for $\lambda$ and various ket-vectors:
\beq
\langle c| = \ck^\dagger \,, \qquad \langle \cb|= - |\cb\rangle^\dagger\,, \qquad \lambda^\dagger = - \lambda \,.
\eeq

%%%%%%%%%%%%%%%%%%%%%%%%%%%%%%%%%%%%%%%%%%%%%%%%%%%%%%%%%%%%%%%%%%%%%%%%%%%%%%
%%%%%%%%%%%%%%%%%%%%%%%%%%%%%%%%%%%%%%%%%%%%%%%%%%%%%%%%%%%%%%%%%%%%%%%%%%%%%%
\newsection{ \large General form of Lagrangian and gauge symmetries for massless, partial-massless, massive, and conformal fields in $(A)dS_{d+1}$ }\label{general}
%%%%%%%%%%%%%%%%%%%%%%%%%%%%%%%%%%%%%%%%%%%%%%%%%%%%%%%%%%%%%%%%%%%%%%%%%%%%%%
%%%%%%%%%%%%%%%%%%%%%%%%%%%%%%%%%%%%%%%%%%%%%%%%%%%%%%%%%%%%%%%%%%%%%%%%%%%%%%

In this section, we formulate our result concerning the general structure of Lagrangian and gauge transformations for spin-$s$ massless, partial-massless, massive, and conformal fields in $(A)dS_{d+1}$. Namely, we demonstrate that how a knowledge of gauge transformations allows us to fix a gauge invariant Lagrangian for just mentioned fields. Below, in Sec.\ref{contransarb}, by using conformal transformation which maps fields in flat space to fields in (A)dS, we obtain gauge transformation of conformal field in (A)dS.
Using then the gauge transformation of field in (A)dS and our result in this Section we find the gauge invariant Lagrangian of conformal field in (A)dS in Sec.\ref{contransarb}.

\noindent {\bf Field content}. In order to formulate our result, it is convenient to use language of generating functions. We note that, by using the oscillators  $\alpha^A$, $\zeta$, and various other oscillators, fields appearing in gauge invariant formulation of massless, partial-massless, massive, and conformal fields can be collected into a ket-vector $\phik$ defined by
\beq
\label{05042014-07} && \phik = \sum_{s'=0}^s \frac{\zeta^{s-s'}}{\sqrt{(s-s')!}} |\phi^{s'}\rangle \,,
\eeq
where the ket-vector $\phik$ should satisfy the constraints
\beq
\label{05042014-08} && (N_\alphabf + N_\zeta)\phik =  s\phik\,,
\\
\label{05042014-10} && (\bar\alphabf^2)^2 \phik = 0\,.
\eeq
Ket-vectors $|\phi^{s'}\rangle$ \rf{05042014-07} depend on the oscillators $\alpha^A$ and various other oscillators. Note however that the ket-vectors $|\phi^{s'}\rangle$ are independent of the oscillator $\zeta$. Constraint  \rf{05042014-08} tells us that the ket-vector $\phik$ should be
degree-$s$ homogeneous polynomial in the oscillators $\alpha^A$, $\zeta$. Constraint \rf{05042014-10} implies that we are using double-traceless tensor fields. Obviously, for the case of $s>1$, we assume the relation $\bar\alphabf^2\phik \ne 0$.

\noindent {\bf Gauge symmetries}. To discuss gauge symmetries we should introduce a set of gauge transformation parameters which we use for a formulation of gauge transformations. We note that, using the oscillators  $\alpha^A$, $\zeta$, gauge transformation parameters appearing in gauge invariant formulation of massless, partial-massless, massive, and conformal fields can be collected into a ket-vector $\xik$ defined by
\beq
\label{05042014-11} && \xik = \sum_{s'=0}^{s-1} \frac{\zeta^{s-1-s'}}{\sqrt{(s-1-s')!}} |\xi^{s'}\rangle \,,
\eeq
where the ket-vector $\xik$ satisfies the constraints
\beq
\label{05042014-12} && (N_\alphabf + N_\zeta)\xik =  (s-1)\xik\,,
\\
\label{05042014-15} && \bar\alphabf^2 \xik = 0\,.
\eeq
Ket-vectors $|\xi^{s'}\rangle$ \rf{05042014-11} depend on the oscillators $\alpha^A$ and various other oscillators. Note however that the ket-vectors $|\xi^{s'}\rangle$ are independent of the oscillator $\zeta$. Constraint  \rf{05042014-11} tells us that the ket-vector $\xik$ should be
degree-$(s-1)$ homogeneous polynomial in the oscillators $\alpha^A$, $\zeta$. Constraint \rf{05042014-15} implies that gauge transformation parameters we are using are traceless tensor fields.

In terms of the ket-vectors $\phik$, $\xik$, a general one-derivative gauge transformation for the ket-vector $\phik$ can be presented as
\be \label{05042014-16}
\delta \phik   =   G \xik\,, \qquad  G \equiv \alphabf \Dbf - e_1 - \alphabf^2 \frac{1}{2N_\alphabf +d-1}\eb_1\,,
\ee
where operators $e_1$ and $\eb_1$ depend on the oscillators $\zeta$, $\bar\zeta$ and other oscillators entering ket-vectors $|\phi^{s'}\rangle$ \rf{05042014-07} and $|\xi^{s'}\rangle$ \rf{05042014-11}. In view of constraint \rf{05042014-15}, gauge transformation \rf{05042014-16} respects constraint in \rf{05042014-10}. Also note that, to respect constraint given in \rf{05042014-08}, the operators $e_1$ and $\eb_1$ should satisfy the commutation relations
\be \label{05042014-16b}
[N_\zeta, e_1] = e_1\,, \qquad [N_\zeta, \eb_1] = -\eb_1\,.
\ee
With the restrictions in \rf{05042014-16b}, transformation given in \rf{05042014-16} is the most general gauge transformation of the ket-vector $\phik$ which does not contain
higher than first order terms in derivatives. Now we are ready to formulate our main statement in this Section.

\noindent {\bf Statement}. General Lagrangian which does not contain
higher than second order terms in derivatives and which respects gauge transformation in \rf{05042014-16} is given by
\beq
\label{05042014-17} \LL  & = & \half e \phibr  E \phik \,,
\\
\label{05042014-18} && E  =  (1-\frac{1}{4}\alphabf^2 \bar\alphabf^2) \bigl( \Box_\pAdS + m_1 + m_2 \alphabf^2\bar\alphabf^2 \bigr) - \Lbf \bar\Lbf\,,
\\
\label{05042014-18a} && \bar\Lbf \equiv  \bar\alphabf \Dbf - \half \alphabf \Dbf  \bar\alphabf^2  -
\eb_1\Pibf^\smponetwo + \half e_1 \bar\alphabf^2\,,
\\
\label{05042014-18b} && \Lbf \equiv \alphabf \Dbf  - \half \alphabf^2 \bar\alphabf \Dbf  - e_1 \Pibf^\smponetwo +
\half \eb_1 \alphabf^2\,,
\eeq
where operators $m_1$, $m_2$ are given by
\beq
\label{05042014-19} && m_1  = -\MM^2 + \rho \Bigl( s(s+d-5) -2d+4 - N_\zeta(2s+d-5-N_\zeta)\Bigr) \,,
\\
\label{05042014-20} && m_2 = \rho\,,
\\
\label{05042014-21} && \MM^2 \equiv - \eb_1 e_1 + \frac{2s + d - 1 - 2N_\zeta}{2s + d - 3 - 2N_\zeta} e_1 \eb_1 \,,
\eeq
while the operators $e_1$, $\eb_1$, and $\MM^2$ \rf{05042014-21} should satisfy the following equations:
\beq
\label{05042014-22} && [\MM^2,e_1] =-2\rho e_1 (2s+d-4-2N_\zeta)\,,
\\
\label{05042014-23} && [\MM^2, \eb_1] = 2\rho (2s+d-4-2N_\zeta)\eb_1\,,
\\
\label{05042014-23x} && e_1^\dagger = - \eb_1\,.
\eeq
The statement is proved by direct computation. The computation is straightforward but tedious. We omit the details of the computation. The following remarks are in order.

\noindent {\bf i}) From \rf{05042014-18}-\rf{05042014-21}, we see that all operators appearing in Lagrangian \rf{05042014-17} are expressed in terms of the operators $e_1$, $\eb_1$ which enter gauge transformation in \rf{05042014-16}. In other words, {\it a knowledge of the operators $e_1$ and $\eb_1$ allows us immediately to fix the gauge invariant Lagrangian}.

\noindent {\bf ii}) From \rf{05042014-21}, we see that operator $\MM^2$ is expressed in terms of the operators $e_1$ and $\eb_1$. This implies that equations \rf{05042014-22},\rf{05042014-23} should be considered as restrictions imposed on the operators $e_1$, $\eb_1$. Equations \rf{05042014-22}-\rf{05042014-23x} provide the complete list of restrictions imposed by gauge symmetries.

\noindent {\bf iii}) For massless, partial-massless, and massive fields,  equations \rf{05042014-22}-\rf{05042014-23x} constitute a complete system of equations which allow us to fix the operators  $e_1$, $\eb_1$ uniquely.

\noindent {\bf iv}) For conformal fields, Eqs.\rf{05042014-22}-\rf{05042014-23x} alone do not allow to fix the operators  $e_1$, $\eb_1$ uniquely. In Refs.\cite{Metsaev:2007fq,Metsaev:2007rw}, we have demonstrated that in order to determine uniquely the operators $e_1$, $\eb_1$ for conformal fields in $R^{d,1}$, we should solve, in addition to restrictions imposed by gauge symmetries, the restrictions imposed by conformal $so(d+1,2)$ symmetries. In Sec.\ref{contransarb}, using the operators $e_1$, $\eb_1$ for conformal field in $R^{d,1}$ and conformal transformation which maps field in flat space to field in (A)dS we obtain operators $e_1$, $\eb_1$ for conformal field in $(A)dS_{d+1}$. The knowledge of operators $e_1$, $\eb_1$ for conformal field in $(A)dS_{d+1}$ allow us then to find Lagrangian of conformal field in $(A)dS_{d+1}$  by using relations in \rf{05042014-19}-\rf{05042014-21}.

\noindent {\bf v}) Let us introduce a new operator $m^2$ defined by the relation
\be
\label{05042014-24} m^2 \equiv  \MM^2 + 2 \rho N_\zeta (2s+d-3-N_\zeta)\,.
\ee
Using \rf{05042014-24}, it is easy to check that equations \rf{05042014-22},\rf{05042014-23} can be represented as
\be \label{05042014-25}
[m^2,e_1]= 0 \,, \qquad  [m^2,\eb_1]=0\,.
\ee
Using \rf{05042014-21}, we see that the operator $m^2$ can be presented in terms of the operators $e_1$ and $\eb_1$ as
\be \label{05042014-27}
m^2 = - \eb_1 e_1 + \frac{2s + d - 1 - 2N_\zeta}{2s + d - 3 - 2N_\zeta} e_1 \eb_1 + 2 \rho N_\zeta (2s+d-3-N_\zeta)\,.
\ee
while the operator $m_1$ appearing in $E$ \rf{05042014-18} takes the form
\be \label{05042014-27a}
m_1 = -m^2 +\rho \Bigl( s(s+d-5) -2d+ 4 +N_\zeta(2s+d-1-N_\zeta)\Bigr)\,.
\ee

%%%%%%%%%%%%%%%%%%%%%%%%%%%%%%%%%%%%%%%%%%%%%%%%%%%%%%%%%%%%%%%%%%%%%%%%%
%%%%%%%%%%%%%%%%%%%%%%%%%%%%%%%%%%%%%%%%%%%%%%%%%%%%%%%%%%%%%%%%%%%%%%%%%
\newsection{ \large Conformal fields in $R^{d,1}$ }\label{review}
%%%%%%%%%%%%%%%%%%%%%%%%%%%%%%%%%%%%%%%%%%%%%%%%%%%%%%%%%%%%%%%%%%%%%%%%%
%%%%%%%%%%%%%%%%%%%%%%%%%%%%%%%%%%%%%%%%%%%%%%%%%%%%%%%%%%%%%%%%%%%%%%%%%

In Sec.\ref{transformation}, we derive our Lagrangian for conformal fields in (A)dS by using the ordinary-derivative Lagrangian formulation of conformal fields in $R^{d,1}$ and applying an appropriate conformal transformation which maps conformal fields in $R^{d,1}$ to conformal fields in $(A)dS_{d+1}$. The ordinary-derivative Lagrangian formulation of conformal fields in $R^{d,1}$ was developed in Refs.\cite{Metsaev:2007fq,Metsaev:2007rw}.  In this Section, we review briefly some our results in Refs.\cite{Metsaev:2007fq,Metsaev:2007rw} which, in Sec.\ref{transformation}, we use to construct a map of conformal fields in $R^{d,1}$ to conformal fields in $(A)dS_{d+1}$.

\subsection{ Spin-0 conformal field in $R^{d,1}$} \label{review-scal}

In the framework of ordinary-derivative approach, spin-0 conformal field (scalar field) is  described by $k+1$ scalar fields
\be  \label{12042014-17}
\phi_{k'}\,,\qquad k' \in [k]_2\,, \qquad k - \hbox{ arbitrary positive integer}\,,
\ee
Here and below, the notation $k' \in [n]_2$
implies that $k' =-n,-n+2,-n+4,\ldots,n-4, n-2,n$:
\be
k' \in [n]_2 \quad \Longrightarrow \quad k' =-n,-n+2,-n+4,\ldots,n-4, n-2,n\,.
\ee
Conformal dimensions of the scalar fields $\phi_{k'}$ are given by
\be
\Delta(\phi_{k'}) = \frac{d-1}{2} + k'\,.
\ee
To simplify the presentation we use oscillators $\upsilon^\oplussm$, $\upsilon^\ominussm$ and collect fields into the ket-vector defined by
\beq
\label{12042014-18} && \phik \equiv \sum_{k'\in [k]_2} \frac{1}{ (\frac{k+k'}{2})!}
(\upsilon^\ominussm)^{^{\frac{k+k'}{2}}}
(\upsilon^\oplussm)^{^{\frac{k-k'}{2}}} \, \phi_{k'} |0\rangle\,,
\\
\label{12042014-19} && (N_\upsilon - k ) \phik = 0 \,.
\eeq
Constraint \rf{12042014-19} tells us that ket-vector $\phik$ \rf{12042014-18} is
degree-$k$ homogeneous polynomial in the oscillators $\upsilon^\oplussm$,
$\upsilon^\ominussm$.  As found in Ref.\cite{Metsaev:2007fq}, ordinary-derivative Lagrangian can be presented as
\be \label{12042014-20}
\LL = \half \phibr (\Box_{_{R^{d,1}}} - m_v^2) \phik\,,  \qquad \Box_{_{R^{d,1}}} \equiv \partial^A \partial^A\,, \qquad  m_v^2 \equiv
\upsilon^\oplussm \bar\upsilon^\oplussm\,,\qquad
\ee
$\partial^A\equiv \eta^{AB}\partial/\partial x^B$. The component form of Lagrangian \rf{12042014-20} takes the form
\be \label{12042014-21}
\LL =  \sum_{k' \in [k]_2} \LL_{k'}\,, \qquad
\LL_{k'} = \half \phi_{-k'} \Box_{_{R^{d,1}}} \phi_{k'} - \half \phi_{-k'}\phi_{k'+2}\,.
\ee

%%%%%%%%%%%%%%%%%%%%%%%%%%%%%%%%%%%%%%%%%%%%%%%%%%%%%%%%%%%%%%%%%%%%%%%%%
%%%%%%%%%%%%%%%%%%%%%%%%%%%%%%%%%%%%%%%%%%%%%%%%%%%%%%%%%%%%%%%%%%%%%%%%%
\subsection{ Arbitrary spin conformal fields in $R^{d,1}$} \label{review-arbspin}
%%%%%%%%%%%%%%%%%%%%%%%%%%%%%%%%%%%%%%%%%%%%%%%%%%%%%%%%%%%%%%%%%%%%%%%%%
%%%%%%%%%%%%%%%%%%%%%%%%%%%%%%%%%%%%%%%%%%%%%%%%%%%%%%%%%%%%%%%%%%%%%%%%%

\noindent {\bf Field content}. To discuss the ordinary-derivative gauge invariant formulation of totally symmetric arbitrary spin-$s$ conformal field in $R^{d,1}$, for arbitrary odd $d\geq 3$, we use the following set of scalar, vector, and tensor fields of the Lorentz algebra $so(d,1)$:
\beq
\label{11042014-01} && \phi_{k'}^{A_1\ldots A_{s'}}\,, \hspace{1.5cm}
s'= \left\{\begin{array}{l}
0,1,\ldots,s;\qquad  \hbox{for odd }\  \ d \geq 5;
\\[3pt]
1,2,\ldots,s;\qquad  \hbox{for } \hspace{0.9cm}  d = 3;
\end{array}\right.
\hspace{2cm} k' \in  [k_{s'}]_2\,;\qquad
\\
\label{11042014-01a} && \hspace{3.6cm} k_{s'} \equiv s'+
\frac{d-5}{2}\,.
\eeq
Tensor fields $ \phi_{k'}^{A_1\ldots A_{s'}}$ are totally symmetric and, when $s'\geq 4$, are double-traceless
\be \label{11042014-02}
\phi_{k'}^{AABBA_5\ldots A_{s'}}=0\,, \qquad s'\geq 4\,.
\ee
We note that the conformal dimension of the field $\phi_{k'}^{A_1\ldots
A_{s'}}$ is given by
\be \label{11042014-03}
\Delta(\phi_{k'}^{A_1\ldots A_{s'}}) =
\frac{d-1}{2} + k'\,.
\ee

To illustrate field content \rf{11042014-01}, we use the
shortcut $\phi_{k'}^{s'}$ for the field $\phi_{k'}^{A_1\ldots
A_{s'}}$ and note that, for arbitrary spin-$s$ conformal field in $R^{d,1}$, $d\geq 5$, the  field content in \rf{11042014-01} can be presented as
{\small
\beq
&& \hspace{0.3cm} \hbox{Field content of spin-$s$ conformal field in $R^{d,1}$, for odd $d \geq 5$, \ \ $s$ - \hbox{arbitrary}}
\nonumber\\
&&\hspace{-1cm} \phi_{-k_s}^s \hspace{1cm} \phi_{2-k_s}^s
\hspace{1cm} \ldots \hspace{1cm}\ldots \hspace{1cm} \ldots
\hspace{1cm} \ldots \hspace{1cm}\ldots \hspace{1cm} \phi_{k_s-2}^s
\hspace{1cm}  \phi_{k_s}^s \qquad
\nonumber\\[15pt]
&& \hspace{-0.3cm} \phi_{1-k_s}^{s-1} \hspace{1cm}
\phi_{3-k_s}^{s-1} \hspace{1cm} \ldots \hspace{1cm} \ldots
\hspace{1cm}\ldots \hspace{1cm}\ldots \hspace{1cm}
\phi_{k_s-3}^{s-1} \hspace{1cm} \phi_{k_s-1}^{s-1}
\nonumber\\[14pt]
&& \hspace{0.5cm} \ldots \hspace{1cm}  \ldots \hspace{1cm}  \ldots
\hspace{1cm} \ldots \hspace{1cm} \ldots   \hspace{1cm} \ldots
\hspace{1cm} \ldots \hspace{1cm} \ldots
\\[14pt]
&& \hspace{0.8cm}  \phi_{s-1-k_s}^1 \hspace{1cm} \phi_{s+1-k_s}^1
\hspace{1cm}  \ldots  \hspace{1cm} \ldots \hspace{1cm}
\phi_{k_s-s-1}^1 \hspace{0.8cm} \phi_{k_s-s+1}^1
\nonumber\\[15pt]
&& \hspace{1.8cm} \phi_{s-k_s}^0 \hspace{1.2cm} \phi_{s-k_s+2}^0
\hspace{1.2cm} \ldots \hspace{1cm} \phi_{k_s-s-2}^0 \hspace{1cm}
\phi_{k_s-s}^0
\nonumber
\eeq }

For $R^{3,1}$, the scalar fields do not enter the field content. Namely, for arbitrary spin-$s$ conformal field in $R^{3,1}$, the field content in \rf{11042014-01} can be presented as
{\small
$$ \hbox{Field content of spin-$s$ conformal field in $R^{3,1}$, $s$ - arbitrary }$$
\be
\begin{array}{ccccccccc}
\phi_{1-s}^s & & \phi_{3-s}^s & & \ldots & & \phi_{s-3}^s & &
\phi_{s-1}^s
\\[12pt]
& \phi_{2-s}^{s-1} & & \phi_{4-s}^{s-1} & \ldots & \phi_{s-4}^{s-1}
& & \phi_{s-2}^{s-1} &
\\[12pt]
& & \ldots  &   & \ldots & & \ldots  &  &
\\[12pt]
&  & & \phi_{-1}^2  & & \phi_1^2 & & &
\\[12pt]
&  & &  & \phi_0^1& & & &
\end{array}
\ee }

To simplify the presentation, we use the oscillators  $\alpha^A$, $\zeta$,
$\upsilon^\oplussm$, $\upsilon^\ominussm$, and collect fields
\rf{11042014-01} into ket-vector $\phik$ defined by
\beq
&& \phik \equiv \sum_{s'=0}^s
\frac{\zeta^{s-s'}}{\sqrt{(s-s')!}}|\phi^{s'}\rangle \,,
\hspace{3cm} \hbox{for } \ R^{d,1}\,, d \geq 5\,,
\nonumber\\[-10pt]
\label{11042014-04} &&
\\[-10pt]
&& \phik \equiv \sum_{s'=1}^s
\frac{\zeta^{s-s'}}{\sqrt{(s-s')!}}|\phi^{s'}\rangle \,,
\hspace{3cm} \hbox{for } \ R^{3,1} \,,
\nonumber\\
\label{11042014-05} && |\phi^{s'}\rangle \equiv   \sum_{k'\in
[k_{s'}]_2} \frac{1}{s'!(\frac{k_{s'} + k'}{2})!}\alpha^{A_1} \ldots
\alpha^{A_{s'}} (\upsilon^\ominussm)^{^{\frac{k_{s'}+k'}{2}}}
(\upsilon^\oplussm)^{^{\frac{k_{s'} - k'}{2}}} \,
\phi_{k'}^{A_1\ldots A_{s'}}|0\rangle\,.
\eeq
From \rf{11042014-02},\rf{11042014-04},\rf{11042014-05},  we see that the ket-vector
$\phik$ satisfies the relations
\beq
\label{11042014-06} && (N_\alphabf + N_\zeta - s)\phik = 0 \,, \hspace{2cm} (N_\zeta + N_\upsilon - k_s) \phik = 0 \,,
\\
\label{11042014-07} && (\bar{\alphabf}^2)^2 \phik = 0 \,,
\eeq
where $k_s$ is defined as in \rf{11042014-01a}.
Relations \rf{11042014-06} tell us that the ket-vector $\phik$ is
degree-$s$ homogeneous polynomial in the oscillators $\alpha^A$,
$\zeta$ and degree-$k_s$ homogeneous polynomial in the oscillators
$\zeta$, $\upsilon^\oplussm$, $\upsilon^\ominussm$. Constraint \rf{11042014-07} is just the presentation of the double-tracelessness  constraints \rf{11042014-02} in terms of the ket-vector $\phik$.

Gauge invariant Lagrangian found in Ref.\cite{Metsaev:2007rw} takes the form
\beq
\label{11042014-08} && \LL = \frac{1}{2} \phibr \mubf (\Box_{_{R^{d,1}}} - m_v^2) \phik + \half \langle \Lb \phi|\Lb\phi\rangle \,,\qquad \Box_{_{R^{d,1}}} \equiv \partial^A \partial^A\,, \qquad  m_v^2 \equiv
\upsilon^\oplussm \bar\upsilon^\oplussm\,,\qquad
\\
\label{11042014-09} && \hspace{1cm} \Lb \equiv  \bar\alphabf \partialbf - \half \alphabf \partialbf \bar\alphabf^2 -
\eb_1 \Pibf^\smponetwo + \half e_1 \bar\alphabf^2\,,\qquad
\\
\label{11042014-09a} &&  \hspace{1cm} e_1 = \zeta e_\zeta \bar\upsilon^\oplussm\,, \qquad \eb_1 = - \upsilon^\oplussm e_\zeta \bar\zeta\,,
\eeq
$\partial^A\equiv \eta^{AB}\partial/\partial x^B$, where operators $\mubf$, $\Pibf^\smponetwo$, and $e_\zeta$ are given in \rf{13042014-15x}-\rf{13042014-16}.

\noindent {\bf Gauge symmetries of conformal field in $R^{d,1}$}. To discuss gauge symmetries of Lagrangian  \rf{11042014-08}, we introduce the following gauge transformation parameters:
\beq
\label{11042014-10} && \xi_{k'-1}^{A_1\ldots A_{s'}}\,,\hspace{1.5cm}
s'=0,1,\ldots,s-1\,,
\hspace{1.5cm} k' \in  [k_{s'}+1]_2\,,
\eeq
where $k_{s'}$ is given in \rf{11042014-01a}. In \rf{11042014-10}, the gauge transformation parameters are scalar, vector, and tensor fields of the Lorentz algebra $so(d,1)$. Tensor fields $\xi_{k'-1}^{A_1\ldots A_{s'}}$ are totally symmetric and, when $s'\geq 2$, are traceless,
\be  \label{11042014-11}
\xi_{k'-1}^{BBA_3\ldots A_{s'}}=0\,, \qquad s'\geq 2\,.
\ee
Conformal dimension of the gauge transformation parameter $\xi_{k'-1}^{A_1\ldots
A_{s'}}$ is given by
\be  \label{11042014-12}
\Delta(\xi_{k'-1}^{A_1\ldots A_{s'}}) = \frac{d-1}{2} + k' -1\,.
\ee

Now, as usually, we collect the gauge transformation parameters into
ket-vector $\xik$ defined by
\beq
\label{11042014-14} && \xik \equiv \sum_{s'=0}^{s-1}
\frac{\zeta^{s-1-s'}}{\sqrt{(s-1-s')!}}|\xi^{s'}\rangle \,,
\\
&& |\xi^{s'}\rangle \equiv  \sum_{k'\in
[k_{s'}+1]_2}\frac{1}{s'!(\frac{k_{s'}+1+k'}{2})!}\alpha^{A_1}
\ldots \alpha^{A_{s'}}
(\upsilon^\ominussm)^{^{\frac{k_{s'}+1+k'}{2}}}
(\upsilon^\oplussm)^{^{\frac{k_{s'}+1-k'}{2}}} \,
\xi_{k'-1}^{A_1\ldots A_{s'}} |0\rangle\,.\qquad
\eeq
Ket-vector $\xik$ \rf{11042014-14} satisfies the algebraic
constraints,
\beq
\label{11042014-15} && (N_\alphabf + N_\zeta - s +1 ) \xik=0
\,, \hspace{1.7cm} (N_\zeta + N_\upsilon - k_s ) \xik=0 \,,
\\
\label{11042014-16} && \bar\alphabf^2 \xik=0 \,,
\eeq
where $k_s$ is defined as in \rf{11042014-01a}.
Relations \rf{11042014-15} tell us
that $\xik$ is a degree-$(s-1)$ homogeneous polynomial in the
oscillators $\alpha^A$, $\zeta$ and degree-$k_s$ homogeneous
polynomial in the oscillators $\zeta$, $\upsilon^\oplussm$,
$\upsilon^\ominussm$. Constraint \rf{11042014-16} is a presentation of the tracelessness  constraints \rf{11042014-11} in terms of the ket-vector $\xik$.

Gauge transformations can entirely be written in terms of $\phik$
and $\xik$ and take the form

\be \label{11042014-16a}
\delta \phik = G \xik  \,,
\qquad G \equiv  \alphabf \partialbf - e_1 - \alphabf^2 \frac{1}{2N_\alphabf
+d-1}\eb_1\,,
\ee
where operators $e_1$, $\eb_1$ are defined in \rf{11042014-09a}.

\noindent { \bf Realization of conformal symmetries in $R^{d,1}$}.
The conformal algebra $so(d+1,2)$ considered in basis of the Lorentz algebra $so(d,1)$ consists of translation generators $\Pbf^A$, dilatation generator $\Dbf$, conformal
boost generators $\Kbf^A$, and generators $\Jbf^{AB}$ which span
$so(d,1)$ Lorentz algebra. We assume the following normalization
for commutators of the conformal algebra:
\beq
&& {}[\Dbf,\Pbf^A]=- \Pbf^A\,, \hspace{2.8cm}  {}[\Pbf^A,\Jbf^{BC}]=\eta^{AB}\Pbf^C - \eta^{AC} \Pbf^B \,,
\nonumber\\
\label{09042014-01} && [\Dbf,\Kbf^A]=\Kbf^A\,, \hspace{3.1cm} [\Kbf^A,\Jbf^{BC}]=\eta^{AB}\Kbf^C -
\eta^{AC} \Kbf^B\,,
\\
&& {}[\Pbf^A, \Kbf^B] = \eta^{AB} \Dbf - \Jbf^{AB}\,, \hspace{1.2cm}  [\Jbf^{AB},\Jbf^{CE}]=\eta^{BC} \Jbf^{AE} + 3\hbox{ terms} \,,
\nonumber\\
\label{09042014-05} && A,B,C,E=0,1,\ldots, d\,, \qquad \quad \eta^{AB} = (-,+,\ldots,+)\,.
\eeq

Let $\phik$ denotes a free conformal field propagating in $R^{d,1}$, $d\geq 3$. Let a Lagrangian for the field $\phik$ be conformal invariant. This implies that the Lagrangian is invariant under transformation (invariance of the Lagrangian is assumed
to be up to total derivatives)
\be
\delta_{\hat{G}} \phik  = \hat{G} \phik
\,,
\ee
where realization of the generators $\hat{G}$ in
terms of differential operators takes the form
\beq
\label{09042014-06} && \Pbf^A = \partial^A \,, \quad \qquad  \Jbf^{AB} = x^A\partial^B -  x^B\partial^A +
M^{AB}\,,
\\
\label{09042014-08} && \Dbf = x^B\partial^B  + \Delta\,,
\\
\label{09042014-09} && \Kbf^A = K_{\Delta,M}^A + R^A\,,
\\
\label{09042014-10} && \hspace{1cm} K_{\Delta,M}^A \equiv - \half x^B x^B \partial^A + x^A \Dbf + M^{AB}x^B\,.
\eeq
In \rf{09042014-08}, $\Delta$ is operator of conformal dimension, while $M^{AB}$ appearing in \rf{09042014-06},\rf{09042014-10} is a spin operator of the Lorentz
algebra $so(d,1)$. Operator $R^A$ appearing in \rf{09042014-09} depends on space-time derivatives $\partial^A \equiv \eta^{AB}\partial/\partial x^B$,  and does not depend on space-time coordinates $x^A$, $[P^A,R^B]=0$. Thus we see that in order to find a realization of conformal symmetries we should fix the operators $\Delta$, $M^{AB}$, and $R^A$.
Realization of these operators on the space of ket-vector $\phik$ \rf{11042014-04} is given by (for scalar field, $\phik$ is given in \rf{12042014-18})
\beq
\label{08062014-01} && M^{AB} = \alpha^A \bar\alpha^B - \alpha^B \bar\alpha^A \,,
\\
\label{08062014-02} && \Delta  =  \frac{d-1}{2}+\Delta'\,,
\qquad \Delta' \equiv  N_{\upsilon^\ominussm} - N_{\upsilon^\oplussm}\,,
\\
&& R^A  = R_\smzero^A + R_\smone^A\,,
\\
&& \hspace{1cm} R_\smzero^A  =   r_{0,1} \bar\alpha^A +  \Awt^A \rb_{0,1}\,, \qquad R_\smone^A =  r_{1,1} \partial^A\,,
\\
\label{14042014-03} && \hspace{1cm} r_{0,1} = 2 \zeta e_\zeta \bar\upsilon^\ominussm\,, \qquad \rb_{0,1} = - 2 \upsilon^\ominussm e_\zeta \bar\zeta\,, \qquad r_{1,1} = -2\upsilon^\ominussm \bar\upsilon^\ominussm \,,
\eeq
where operators $\Awt^A$ and $e_\zeta$ are given in \rf{13042014-15} and \rf{13042014-16} respectively. Note that, for scalar field, we get  $M^{AB}=0$, $R^A = r_{1,1}\partial^A$, and $\Delta$ takes the same form as in \rf{08062014-02}.

\noindent {\bf so(d+1,2) algebra in bases of $so(d)$ and $so(d-1,1)$ subalgebras}. In order to relate conformal fields in $R^{d,1}$ to the ones in $(A)dS_{d+1}$, we use the Poincar\'e  parametrization of $(A)dS_{d+1}$ given by
\beq
\label{10042014-03} && ds^2 = \frac{R^2}{z^2}(\eta_{ab} dx^a dx^b - \epsilon dz dz)\,,
\\
\label{10042014-03b1} && \eta^{ab} = (+,+,\ldots,+)\,, \qquad \epsilon =1\,,\ \ \ \qquad a,b =1,2,\ldots, d\,, \hspace{1.7cm} \hbox{ for dS}
\\
\label{10042014-03b2} && \eta^{ab} = (-,+,\ldots,+)\,, \qquad \epsilon =-1\,, \qquad a,b=0,1,\ldots, d-1\,, \hspace{1cm} \hbox{ for AdS}\qquad
\eeq
Manifest symmetries of line element \rf{10042014-03} are described, for the case of dS, by $so(d)$ algebra and, for the case of AdS, by $so(d-1,1)$ algebra. Therefore it is reasonable to represent conformal symmetries in the bases of the respective $so(d)$ and $so(d-1,1)$ algebras. To this end we note that the Cartesian coordinates $x^A$ in $R^{d,1}$ can be related to the Poincar\'e coordinates $z$, $x^a$ in \rf{10042014-03} by using the following identification of the radial coordinate $z$ and Cartesian coordinates $x^0$ and $x^d$:
\beq
&& x^0 \equiv z, \hspace{2cm} \hbox{ for dS}
\nonumber\\[-10pt]
&& \label{10042014-04}
\\[-10pt]
&& x^d \equiv z, \hspace{2cm} \hbox{ for AdS}
\nonumber
\eeq
For dS, the remaining Cartesian coordinates $x^A$, $A=1,2,\ldots, d$, are identified with the Poincar\'e coordinates $x^a$, $a=1,2,\ldots, d$, while, for AdS, the remaining Cartesian coordinates $x^A$, $A=0,1,\ldots, d-1$, are identified with the Poincar\'e coordinates $x^a$, $a=0,1,2,\ldots, d-1$. In other words, taking into account the identifications in \rf{10042014-04}, we use the following spitting of the Cartesian  coordinates $x^A$ into the Poincar\'e coordinates $z$, $x^a$:
\beq
&& x^A = z, x^a\,,  \qquad a=1,2,\ldots, d\,, \hspace{1.7cm} \hbox{ for dS}
\nonumber\\[-10pt]
&& \label{10042014-05}
\\[-10pt]
&& x^A = z, x^a\,, \qquad a=0,1,\ldots, d-1\,, \hspace{1cm} \hbox{ for AdS}
\nonumber
\eeq
We note that the decomposition of the coordinate given in \rf{10042014-05} implies the following decomposition of flat metric tensor and scalar products:
\beq
\label{12042014-24} && \eta^{AB} = \eta^{zz}, \eta^{ab}
\\
\label{12042014-25} && \eta^{zz} = - \epsilon \,, \qquad \eta^{ab} = (\epsilon, +,\ldots, +)
\\
&& \eta_{AB} X^A Y^B = -\epsilon X^z Y^z +  X^a Y^a \,,\qquad X^aY^a \equiv \eta_{ab} X^a Y^b \,.
\eeq
Also, the decomposition of coordinates in \rf{10042014-05} implies the following decomposition of generators of the conformal algebra $so(d+1,2)$ into generators of (A)dS space isometry symmetries and generators of (A)dS space conformal boost symmetries
\beq
\label{10042014-01} && \Pbf^a\,, \quad \Dbf\,,\quad \Kbf^a\,, \quad \Jbf^{ab} \hspace{1cm}  \hbox{ (A)dS$_{d+1}$ space isometry symmetries }
\\
\label{10042014-02} && \Pbf^z\,,\quad \Kbf^z\,,\quad \Jbf^{za},  \hspace{2cm} \hbox{ (A)dS$_{d+1}$ space conformal boost symmetries }\qquad
\eeq
We note that, for dS, generators $\Jbf^{ab}$ in \rf{10042014-01} span $so(d)$ algebra, while, for AdS, generators $\Jbf^{ab}$ in \rf{10042014-01} span $so(d-1,1)$ algebra.  Realization of generators \rf{10042014-01},\rf{10042014-02} in terms of differential operators is obtained from \rf{09042014-06}-\rf{09042014-10},
\beq
\label{11042014-17} && \Pbf^a = \partial^a \,, \qquad  \Jbf^{ab} = x^a\partial^b -  x^b\partial^a +
M^{ab}\,,
\\
\label{11042014-19} && \Dbf = x^a\partial^a + z\partial_z  + \Delta\,,
\\
\label{11042014-20} && \Kbf^a = K_{\Delta,M}^a + R^a\,,
\\
\label{11042014-21} && \hspace{1cm} K_{\Delta,M}^a \equiv - \half (x^b x^b -\epsilon z^2) \partial^a + x^a \Dbf + M^{ab}x^b + \epsilon M^{za} z\,,
\eeq
\beq
&& \Pbf^z = - \epsilon \partial_z \,,
\\
&& \Jbf^{za} = z\partial^a + \epsilon x^a\partial_z +
M^{za}\,,
\\
&& \Kbf^z = K_{\Delta,M}^z + R^z\,,
\\
&& \hspace{1cm} K_{\Delta,M}^z \equiv  \half ( \epsilon x^b x^b - z^2)\partial_z + z \Dbf + M^{za}x^a\,,
\eeq
where expressions for the spin operators  $M^{AB}=M^{za},M^{ab}$, the conformal dimension operator $\Delta$, and operator $R^A=R^z,R^a$ can be read from \rf{08062014-01}-\rf{14042014-03}.

%%%%%%%%%%%%%%%%%%%%%%%%%%%%%%%%%%%%%%%%%%%%%%%%%%%%%%%%%%%%%%%%%%%%%%%%%
%%%%%%%%%%%%%%%%%%%%%%%%%%%%%%%%%%%%%%%%%%%%%%%%%%%%%%%%%%%%%%%%%%%%%%%%%
\subsection{ Relativistic symmetries of fields in $(A)dS_{d+1}$}
%%%%%%%%%%%%%%%%%%%%%%%%%%%%%%%%%%%%%%%%%%%%%%%%%%%%%%%%%%%%%%%%%%%%%%%%%
%%%%%%%%%%%%%%%%%%%%%%%%%%%%%%%%%%%%%%%%%%%%%%%%%%%%%%%%%%%%%%%%%%%%%%%%%

Relativistic symmetries of fields in $(A)dS_{d+1}$ are described by the $so(d+1,1)$ algebra for the case of dS and by $so(d,2)$ algebra for the case of AdS. For the description of field dynamics in  $(A)dS_{d+1}$, we used tensor fields of $so(d,1)$ algebra. However, as we prefer to realize the algebra of $(A)dS_{d+1}$ space symmetries as subalgebra of conformal symmetries considered, for the case of dS, in the basis of $so(d)$ algebra and, for the case of AdS, in the basis of $so(d-1,1)$ algebra (see \rf{10042014-01}), it is reasonable to represent the  $so(d+1,1)$ and $so(d,2)$ algebras in the respective bases of $so(d)$ and $so(d-1,1)$ algebras. We recall that $so(d+1,1)$ and $so(d,2)$ algebras considered in the respective bases of $so(d)$ and $so(d-1,1)$ algebras consist of translation generators $P^a$, conformal boost generators $K^a$, dilatation generator $D$, and generators of the respective $so(d)$ and $so(d-1,1)$ algebras, $J^{ab}$ . Commutators of generators of $so(d+1,1)$ and $so(d,2)$ algebras take the form
\beq
&& {}[D,P^a]=- P^a\,, \hspace{2.3cm}  {}[P^a,J^{bc}]=\eta^{ab}P^c
-\eta^{ac} P^b \,,
\nonumber\\
&& [D,K^a]=K^a\,, \hspace{2.5cm} [K^a,J^{bc}]=\eta^{ab}K^c -
\eta^{ac} K^b\,,
\\
&& {}[P^a, K^b] = \eta^{ab} D - J^{ab}\,, \hspace{1cm}  [J^{ab},J^{ce}]=\eta^{bc} J^{ae} + 3\hbox{ terms} \,,
\nonumber
\eeq
where $\eta^{ab}$ is given in \rf{10042014-03b1},\rf{10042014-03b2}.
Realization of the generators in terms of differential operators acting on field propagating in (A)dS is well-known,
\beq
\label{11042014-22} && P^a = \partial^a \,,\quad\qquad  J^{ab} = x^a\partial^b -  x^b\partial^a + M^{ab}\,,
\\
\label{11042014-24} && D = x^a\partial^a + z\partial_z\,,
\\
\label{11042014-25} && K^a = -\frac{1}{2}(x^b x^b - \epsilon z^2) \partial^a + x^a D + M^{ab} x^b + \epsilon  M^{za} z\,,
\eeq
where $M^{AB}= M^{za}, M^{ab}$ is spin operator of the Lorentz algebra $so(d,1)$,
\beq
&& [M^{AB},M^{CE}]=\eta^{BC} M^{AE} + 3 \hbox{ terms} \,,
\\
&& M^{AB} = \alpha^A \bar\alpha^B - \alpha^B \bar\alpha^A \,, \qquad [\bar\alpha^A,\alpha^B] = \eta^{AB}\,.
\eeq

%%%%%%%%%%%%%%%%%%%%%%%%%%%%%%%%%%%%%%%%%%%%%%%%%%%%%%%%%%%%%%%%%%%%%%%%%
%%%%%%%%%%%%%%%%%%%%%%%%%%%%%%%%%%%%%%%%%%%%%%%%%%%%%%%%%%%%%%%%%%%%%%%%%
\newsection{ \large Conformal transformation from fields in $R^{d,1}$ to fields in $(A)dS_{d+1}$ }\label{transformation}
%%%%%%%%%%%%%%%%%%%%%%%%%%%%%%%%%%%%%%%%%%%%%%%%%%%%%%%%%%%%%%%%%%%%%%%%%
%%%%%%%%%%%%%%%%%%%%%%%%%%%%%%%%%%%%%%%%%%%%%%%%%%%%%%%%%%%%%%%%%%%%%%%%%

In this Section, we demonstrate our method for the derivation of Lagrangian for conformal field in (A)dS. For scalar field, using the formulation of conformal field in $R^{d,1}$ described in Sec.\ref{review-scal} and applying conformal transformation which maps conformal field in $R^{d,1}$ to conformal field in $(A)dS_{d+1}$, we obtain Lagrangian of scalar conformal field in $(A)dS_{d+1}$. For arbitrary spin field, using the gauge transformation rule of conformal field in $R^{d,1}$ described in Sec.\ref{review-arbspin} and applying conformal transformation which maps the conformal field in $R^{d,1}$ to conformal field in $(A)dS_{d+1}$, we obtain gauge transformation rule of the arbitrary spin conformal field in $(A)dS_{d+1}$. Using then the gauge transformation rule of the arbitrary spin conformal field in (A)dS and our result in Sec.\ref{general}, we find the gauge invariant Lagrangian of the arbitrary spin conformal field in (A)dS. We consider the scalar and arbitrary spin conformal fields in turn.

%%%%%%%%%%%%%%%%%%%%%%%%%%%%%%%%%%%%%%%%%%%%%%%%%%%%%%%%%%%%%%%%%%%%%%%%%
%%%%%%%%%%%%%%%%%%%%%%%%%%%%%%%%%%%%%%%%%%%%%%%%%%%%%%%%%%%%%%%%%%%%%%%%%
\subsection{ Conformal transformation for scalar field }
%%%%%%%%%%%%%%%%%%%%%%%%%%%%%%%%%%%%%%%%%%%%%%%%%%%%%%%%%%%%%%%%%%%%%%%%%
%%%%%%%%%%%%%%%%%%%%%%%%%%%%%%%%%%%%%%%%%%%%%%%%%%%%%%%%%%%%%%%%%%%%%%%%%

In this Section, we discuss conformal transformation which maps scalar field in $R^{d,1}$ to the one in $(A)dS_{d+1}$. It is convenient to realize the conformal transformation in two steps. We now discuss these steps in turn.

\noindent {\bf Step 1. Derivation of intermediate form of Lagrangian for conformal field in (A)dS}. The conformal transformation which allows us to find intermediate Lagrangian for conformal field in (A)dS is found by matching generators of relativistic symmetries for
(A)dS fields given in \rf{11042014-22}-\rf{11042014-25} and the ones for fields in flat space given in \rf{11042014-17}-\rf{11042014-20}. Let us use the notation $|\phi_{{R^{d,1}}}\rangle$ and $|\phi_{{\pAdS_{d+1}}}^\intm\rangle$ for the respective ket-vectors of scalar fields in $R^{d,1}$ and $(A)dS_{d+1}$.  The ket-vectors are related by the conformal transformation given by
\be \label{11042014-26}
|\phi_{_{R^{d,1}}}\rangle = U^\intm |\phi_{_{\pAdS_{d+1}}}^\intm \rangle\,,
\ee
where $U^\intm$ stands for operator of conformal transformation. This operator is found by matching the generators in \rf{11042014-17}-\rf{11042014-20} and the respective generators in \rf{11042014-22}-\rf{11042014-25}. In other words, the operator $U^\intm$ is found by solving the following equations:
\beq
\label{11042014-27a} && \Pbf^a |\phi_{_{R^{d,1}}}\rangle = U^\intm P^a |\phi_{_{\pAdS_{d+1}}}^\intm \rangle\,,
\\
\label{11042014-28b} && \Jbf^{ab} |\phi_{_{R^{d,1}}}\rangle = U^\intm J^{ab} |\phi_{_{\pAdS_{d+1}}}^\intm \rangle\,,
\\
\label{11042014-27} && \Dbf |\phi_{_{R^{d,1}}}\rangle = U^\intm D |\phi_{_{\pAdS_{d+1}}}^\intm \rangle\,,
\\
\label{11042014-28} && \Kbf^a |\phi_{_{R^{d,1}}}\rangle = U^\intm K^a |\phi_{_{\pAdS_{d+1}}}^\intm \rangle\,.
\eeq
Comparing \rf{11042014-17} with \rf{11042014-22}, we see that Eqs.\rf{11042014-27a},\rf{11042014-28b} are already satisfied. All that remains is to solve Eqs.\rf{11042014-27},\rf{11042014-28}. Solution to Eqs.\rf{11042014-27},\rf{11042014-28} is found to be
\beq
\label{12042014-01} U^\intm & = &  |\rho|^{\frac{1-d}{4}} \exp(-\frac{\epsilon\tau}{2} X)  z^{-\Delta}\,,
\\
\label{12042014-02} && \tau \equiv 4 \upsilon^\ominussm \bar\upsilon^\ominussm \,, \qquad X \equiv \half \{ \frac{1}{z},\partial_z\}\,.
\eeq
Plugging \rf{11042014-26},\rf{12042014-01} into Lagrangian for conformal field in flat space \rf{12042014-20}, we get Lagrangian for conformal field in (A)dS,
\beq
\label{12042014-03} \LL  & = &  \half (|\rho|z^2)^{\frac{1-d}{2}} \langle \phi_{_{\pAdS_{d+1}}}^\intm |\Bigl( \Box - \epsilon (\partial_z^2 + \frac{1-d}{z}\partial_z)- \frac{1}{z^2}(\nu^2 + \frac{\epsilon d^2}{4}) \Bigr) |\phi_{_{\pAdS_{d+1}}}^\intm \rangle\,,
\nonumber\\
&& \Box \equiv \partial^a \partial^a\,, \qquad \nu^2 \equiv  V^\oplussm \Vb^\oplussm - \epsilon (N_v + \half)^2 \,.
\eeq
We now recall that, in the Poincar\'e coordinates \rf{10042014-03}, the D'Alembert operator for scalar field in $(A)dS_{d+1}$ takes the form
\be \label{12042014-07x}
\DD^2 =  |\rho| z^2\Bigl(\Box -\epsilon (\partial_z^2 + \frac{1-d}{z}\partial_z )\Bigr)\,.
\ee
Using \rf{12042014-07x}, we see that Lagrangian \rf{12042014-03} is the presentation of the following covariant Lagrangian in terms of Poincar\'e parametrization of (A)dS \rf{10042014-03},
\beq
\label{12042014-07} \LL &  = & \half e \langle \phi_{_{\pAdS_{d+1}}}^\intm |\bigl( \DD^2 - m_\intm^2 \bigl) |\phi_{_{\pAdS_{d+1}}}^\intm \rangle\,,
\\
\label{12042014-08} && m_\intm^2 \equiv  |\rho| V^\oplussm \Vb^\oplussm +   \rho \bigl( \frac{d^2}{4} - (N_v + \half)^2 \bigr) \,,
\eeq
where we  use the ``deformed oscillators" $V^\oplussm$, $\Vb^\oplussm$ defined by the relations
\be \label{12042014-09}
V^\oplussm = (1+ \epsilon\tau)^{1/4} \upsilon^\oplussm (1+\epsilon\tau)^{1/4}\,,
\qquad  \Vb^\oplussm = (1+\epsilon\tau)^{1/4}\bar\upsilon^\oplussm  (1+\epsilon\tau)^{1/4}\,,
\ee
which can also be represented as
\be \label{12042014-10}
V^\oplussm \equiv \half \{\sqrt{1+\epsilon\tau}, \upsilon^\oplussm\}\,, \qquad
\Vb^\oplussm \equiv \half \{\sqrt{1+\epsilon\tau}, \bar\upsilon^\oplussm\}\,.
\ee
We refer to Lagrangian \rf{12042014-07} as intermediate Lagrangian. The intermediate Lagrangian describes conformal field in arbitrary parametrization of (A)dS space. Note that mass operator $m_\intm^2$ \rf{12042014-08} entering the intermediate Lagrangian is not diagonal on space of ket-vector $|\phi_{{\pAdS_{d+1}}}^\intm \rangle$.
The mass operator can be diagonalized by an appropriate transformation of $|\phi_{{\pAdS_{d+1}}}^\intm \rangle$.
To this end we proceed to Step 2 of our procedure.

\noindent {\bf Step 2. Derivation of factorized form of Lagrangian for conformal field in (A)dS}. In order to diagonalize the mass operator $m_\intm^2$ \rf{12042014-08} we make the following transformation:%
\footnote{ Note that equations \rf{11042014-27a}-\rf{11042014-28} do not fix the operator $U^\intm$ uniquely. This is to say that the $U^\intm$ is defined up to unitary operator that is independent of space-time coordinates and derivatives.}
\beq
\label{12042014-11} && |\phi_{_{\pAdS_{d+1}}} \rangle = \Uwh |\phi_{_{\pAdS_{d+1}}}^\intm \rangle\,, \qquad   \Uwh \Uwh^\dagger = 1 \,,
\\
\label{12042014-12} \Uwh & = & \sum_{l,n=0}^k u_{ln}\vartheta^l \chi^{k-l}|0\rangle\langle0| (\bar\upsilon^\ominussm)^k (V^\oplussm)^n (\Vb^\oplussm)^n\,,
\nonumber\\
&& u_{ln} =  \frac{(-\epsilon)^n(l-n+k)!}{(l+n-k)!}u_l\,, \hspace{1.1cm}  \hbox{ for } \ \ l+n \geq k\,,
\nonumber\\
&& u_{ln} = 0\,, \hspace{4.4cm} \hbox{ for } \ \ l+n <  k\,,
\nonumber\\
&& u_l = \frac{(-)^{l-k} }{k!(k-l)!} \Bigl( \frac{2l+1}{l!(k+l+1)!}\Bigr)^{1/2}\,.
\eeq
Note that the intermediate ket-vector $|\phi_{_{\pAdS_{d+1}}}^\intm\rangle$ defined in \rf{11042014-26} depends on the oscillator $\upsilon^\oplussm$, $\upsilon^\ominussm$, while the new ket-vector $|\phi_{_{\pAdS_{d+1}}}\rangle$ defined in \rf{12042014-11} depends on new oscillators $\vartheta$, $\chi$. For our oscillator algebra, see Appendix. In terms of the new ket-vector $|\phi_{_{\pAdS_{d+1}}}\rangle$, the intermediate Lagrangian given in \rf{12042014-07} takes the same form as in \rf{12042014-07} with the desired diagonalized mass operator,
\beq
\label{12042014-15} \LL &  = & \half e \langle \phi_{_{\pAdS_{d+1}}} |\bigl( \DD^2 - m_\diag^2 \bigl) |\phi_{_{\pAdS_{d+1}}}\rangle\,,
\\
\label{12042014-16}  && m_\diag^2 \equiv  \rho \bigl( \frac{d^2}{4} - (N_\vartheta + \half)^2 \bigr) \,.
\eeq
In terms of $|\phi_{_{\pAdS_{d+1}}}\rangle$  \rf{12042014-11}, constraint \rf{12042014-19} takes the form
\be \label{13042014-30}
(N_\vartheta + N_\chi - k) |\phi_{_{\pAdS_{d+1}}}\rangle = 0 \,.
\ee
Constraint \rf{13042014-30} implies that the ket-vectors  $|\phi_{_{\pAdS_{d+1}}}\rangle$ is decomposed into oscillators as
\be \label{13042014-31}
|\phi_{_{\pAdS_{d+1}}}\rangle = \sum_{k'=0}^k \frac{\vartheta^{k-k'}\chi^{k'}}{\sqrt{k'!(k-k')!}} \phi_{k'}|0\rangle\,.
\ee
Plugging \rf{13042014-31} into \rf{12042014-15} and using \rf{13042014-32}, we get component form of Lagrangian given in \rf{07042014-02}.

%%%%%%%%%%%%%%%%%%%%%%%%%%%%%%%%%%%%%%%%%%%%%%%%%%%%%%%%%%%%%%%%%%%%%%%%%
%%%%%%%%%%%%%%%%%%%%%%%%%%%%%%%%%%%%%%%%%%%%%%%%%%%%%%%%%%%%%%%%%%%%%%%%%
\subsection{ Conformal transformation for arbitrary spin field }\label{contransarb}
%%%%%%%%%%%%%%%%%%%%%%%%%%%%%%%%%%%%%%%%%%%%%%%%%%%%%%%%%%%%%%%%%%%%%%%%%
%%%%%%%%%%%%%%%%%%%%%%%%%%%%%%%%%%%%%%%%%%%%%%%%%%%%%%%%%%%%%%%%%%%%%%%%%

In this Section, we discuss conformal transformation which maps arbitrary spin conformal field in $R^{d,1}$ to the one in $(A)dS_{d+1}$. As in the case of scalar field, it is convenient to realize the conformal transformation in two steps. We now discuss these steps in turn.

\noindent {\bf Step 1. Derivation of intermediate form of Lagrangian for conformal field in (A)dS}. The conformal transformation which allows us to find intermediate Lagrangian for conformal field in (A)dS is found by matching generators of relativistic symmetries for
fields in (A)dS given in \rf{11042014-22}-\rf{11042014-25} and the ones for fields in flat space given in \rf{11042014-17}-\rf{11042014-20}. Using the notation $|\phi_{{R^{d,1}}}\rangle$ and $|\phi_{{\pAdS_{d+1}}}^\intm\rangle$ for the respective ket-vectors of fields in $R^{d,1}$ and $(A)dS_{d+1}$, we consider the conformal transformation given by
\be \label{13042014-33a}
|\phi_{_{R^{d,1}}}\rangle = U^\intm |\phi_{_{\pAdS_{d+1}}}^\intm \rangle\,,
\ee
where $U^\intm$ stands for the operator of conformal transformation.
This operator is found by matching the generators in \rf{11042014-17}-\rf{11042014-20} and the respective generators in \rf{11042014-22}-\rf{11042014-25}. In other words, the operator $U^\intm$ is found by solving the equations given in \rf{11042014-27a}-\rf{11042014-28}.
Comparing \rf{11042014-17} with \rf{11042014-22}, we see that Eqs.\rf{11042014-27a},\rf{11042014-28b} are already satisfied. All that remains is to solve Eqs.\rf{11042014-27},\rf{11042014-28}. Solution to Eqs.\rf{11042014-27},\rf{11042014-28} is found to be
\beq
\label{13042014-33} U^\intm & = &  U_1 U_0 \,,
\\
\label{13042014-34} U_1 & \equiv  &  |\rho|^{\frac{1-d}{4}} \exp(-\frac{\epsilon\tau}{2} X)  z^{-\Delta} \,,
\\
\label{13042014-35} U_0 & \equiv &  \exp( -t_c R_\smzero^z ) \,,
\\
\label{12042014-36} && \tau \equiv 4 \upsilon^\ominussm \bar\upsilon^\ominussm \,, \qquad X \equiv \half \{ \frac{1}{z},\partial_z\}\,.
\\
&& R_\smzero^z = r_{0,1} \bar\alpha^z + \Awt^z \rb_{0,1}\,, \qquad
\Awt^z  \equiv \alpha^z - \alphabf^2\frac{1}{2s+d-5 - 2N_\zeta} \bar\alpha^z\,,
\\
\label{13042014-37} && \cos\omega t_c = \sqrt{1+\epsilon\tau}\,,\qquad  \frac{\sin\omega t_c}{\omega}=1\,, \qquad \omega^2 \equiv -\epsilon \tau\,,
\eeq
where relations \rf{13042014-37} are considered as a definition of $t_c$. For definition of $r_{0,1}$, $\rb_{0,1}$, see \rf{14042014-03}.
Note that, in terms of $|\phi_{_{\pAdS_{d+1}}}^\intm \rangle$, constraints \rf{11042014-06},\rf{11042014-07} are represented as
\be \label{13042014-43}
(N_\alphabf + N_\zeta - s)|\phi_{_{\pAdS_{d+1}}}^\intm \rangle= 0\,, \hspace{0.5cm}
(N_\zeta + N_\upsilon  - k_s) |\phi_{_{\pAdS_{d+1}}}^\intm \rangle = 0\,, \qquad (\bar\alphabf^2)^2|\phi_{_{\pAdS_{d+1}}}^\intm \rangle = 0 \,.
\ee

Now our purpose is to find how gauge transformation of $|\phi_{R^{d,1}} \rangle$  given in \rf{11042014-16a} is realized on space $|\phi_{_{\pAdS_{d+1}}}^\intm \rangle$. To this end, adopting the notation $|\xi_{_{R^{d,1}}}\rangle$ for the gauge transformation parameter $\xik$ appearing in \rf{11042014-14},\rf{11042014-16a}, we make the following conformal transformation:
\be \label{13042014-38}
|\xi_{_{R^{d,1}}}\rangle = U^\intm |\rho|^{1/2} z |\xi_{_{\pAdS_{d+1}}}^\intm \rangle\,.
\ee
Using \rf{13042014-33a},\rf{13042014-38} and adopting notation $G_{R^{d,1}}$ for operator $G$ appearing in \rf{11042014-16a}, we find the following relation:
\be \label{13042014-39}
G_{R^{d,1}}|\xi_{_{R^{d,1}}}\rangle = U^\intm |\rho|^{1/2} z G_{_{\pAdS_{d+1}}}^\intm |\xi_{_{\pAdS_{d+1}}}^\intm \rangle\,,
\ee
where a new gauge transformation operator $G_{_{\pAdS_{d+1}}}^\intm$ appearing in \rf{13042014-39} takes the form
\beq
\label{13042014-40} G_{_{\pAdS_{d+1}}}^\intm & = &   \alphabf \Dbf - e_1^\intm - \alphabf^2 \frac{1}{2N_\alphabf +d-1}\eb_1^\intm\,,
\\
\label{13042014-41} && e_1^\intm = \zeta e_\zeta \Vb^\oplussm \,, \qquad \eb_1^\intm = - V^\oplussm e_\zeta \bar\zeta\,,
\eeq
where the ``deformed oscillators" $V^\oplussm$, $\Vb^\oplussm$ are defined in \rf{12042014-09}, while the $e_\zeta$ is given in \rf{13042014-16}.

Thus we see that gauge transformation operator $G_{_{\pAdS_{d+1}}}$ \rf{13042014-40} takes the form we discussed in Sec.\ref{general}. This is to say that a knowledge of the explicit form of operators $e_1$, $\eb_1$ given in \rf{13042014-41} and relations in \rf{05042014-17}-\rf{05042014-18b},\rf{05042014-20},\rf{05042014-27},\rf{05042014-27a} allows us to find explicit form of the gauge invariant Lagrangian. Plugging $e_1$, $\eb_1$ \rf{13042014-41} into \rf{05042014-27}, we get the following two equivalent realizations of the operator $m_\intm^2$ on space of $|\phi_{_{\pAdS_{d+1}}}^\intm\rangle$:
\beq
\label{13042014-42} m_\intm^2  & = &  |\rho| V^\oplussm \Vb^\oplussm  - \rho (N_\upsilon + \half)^2 + \rho \bigl( s+\frac{d-4}{2} \bigr)^2
\nonumber\\
& =& |\rho| V^\oplussm \Vb^\oplussm  + \rho N_\zeta(2s+d-4-N_\zeta)\,.
\eeq
Note that, for the derivation of representations for $m_\intm^2$ in \rf{13042014-42}, one needs to use the second constraint in \rf{13042014-43}. Finally, we verify that  operators $e_1^\intm$, $\eb_1^\intm$ \rf{13042014-41}, $m_\intm^2$ \rf{13042014-42}  satisfy equations \rf{05042014-23x}, \rf{05042014-25}.

To summarize, relations \rf{13042014-40}-\rf{13042014-42} together with the ones in \rf{05042014-17}-\rf{05042014-18b},\rf{05042014-20},\rf{05042014-27a} provide us the description of intermediate Lagrangian for arbitrary spin-$s$ conformal field in $(A)dS_{d+1}$. Note however that mass operator $m_\intm^2$ \rf{13042014-42} entering the intermediate Lagrangian is not diagonal on space of ket-vector $|\phi_{_{\pAdS_{d+1}}}^\intm \rangle$. In other words, the intermediate Lagrangian with mass operator $m_\intm^2$ in \rf{13042014-42} does not provide the factorized description of conformal field in (A)dS.
The mass operator can be diagonalized by an appropriate unitary transformation of $|\phi_{_{\pAdS_{d+1}}}^\intm \rangle$. To this end we proceed to Step 2 of our procedure.

\noindent {\bf Step 2. Derivation of factorized form of Lagrangian for conformal field in (A)dS}. In order to diagonalize the operator $m_\intm^2$ \rf{13042014-42} we make the following transformation:
\beq
\label{13042014-44} && \hspace{-1.5cm} |\phi_{_{\pAdS_{d+1}}} \rangle = \Uwh |\phi_{_{\pAdS_{d+1}}}^\intm \rangle\,, \qquad \qquad \Uwh \Uwh^\dagger = 1 \,,
\\
\label{13042014-45} \Uwh & = & \sum_{s'=0}^s \sum_{l,n=0}^{k_{s'}} u_{ln}^{(s')}\vartheta^l \chi^{k_{s'}-l}\, \Pi_\zeta^{(s-s')}\, (\bar\upsilon^\ominussm)^{k_{s'}} (V^\oplussm)^n (\Vb^\oplussm)^n \,,
\nonumber\\
&& u_{ln}^{(s')} =  \frac{(-\epsilon)^n(l-n+k_{s'})!}{(l+n-k_{s'})!}u_l^{(s')}\,, \hspace{1.1cm}  \hbox{ for } \ \ l+n \geq k_{s'}\,,
\nonumber\\
&& u_{ln}^{(s')} = 0\,, \hspace{5cm} \hbox{ for } \ \ l+n <  k_{s'}\,,
\nonumber\\
&& u_l^{(s')} = \frac{(-)^{l-k_{s'}}}{k_{s'}!(k_{s'}-l)!} \Bigl( \frac{2l+1}{l!(k_{s'}+l+1)!}\Bigr)^{1/2}\,,
\nonumber\\
&& \Pi_\zeta^{(s-s')} \equiv \frac{\zeta^{s-s'}}{\sqrt{(s-s')!}} |0\rangle \langle 0| \frac{\bar\zeta^{s-s'}}{\sqrt{(s-s')!}}\,,
\eeq
where $k_{s'}$ is given in \rf{11042014-01a}. Note that the intermediate ket-vector $|\phi_{{\pAdS_{d+1}}}^\intm\rangle$ defined in \rf{13042014-33a} depends on the oscillators $\alpha^A$, $\zeta$, $\upsilon^\oplussm$, $\upsilon^\ominussm$, while the new ket-vector $|\phi_{{\pAdS_{d+1}}}\rangle$ defined in \rf{13042014-44} depends on oscillators $\alpha^A$, $\zeta$, $\vartheta$, $\chi$. In terms of $|\phi_{{\pAdS_{d+1}}}\rangle$, constraints \rf{13042014-43} take the form
\be \label{13042014-51}
(N_\alphabf + N_\zeta - s)|\phi_{_{\pAdS_{d+1}}}\rangle = 0 \,, \qquad (N_\zeta + N_\vartheta + N_\chi  - k_s) |\phi_{_{\pAdS_{d+1}}}\rangle = 0 \,, \qquad (\bar\alphabf^2)^2|\phi_{_{\pAdS_{d+1}}}\rangle =0 \,.
\ee
Constraints \rf{13042014-51} are easily obtained from the ones in \rf{13042014-43} by using \rf{13042014-44}. Note that the first and second constraints in \rf{13042014-51} imply that $|\phi_{_{\pAdS_{d+1}}}\rangle$ can be represented as in \rf{04042014-19},\rf{04042014-20}.

Now our purpose is to find how gauge transformation of $|\phi_{{\pAdS_{d+1}}}^\intm \rangle$  given in \rf{13042014-40} is realized on space $|\phi_{{\pAdS_{d+1}}}\rangle$ given in \rf{13042014-44}. To this end, we make the following transformation of gauge transformation parameter:
\be \label{13042014-46}
|\xi_{_{\pAdS_{d+1}}} \rangle = \Uwh |\xi_{_{\pAdS_{d+1}}}^\intm \rangle\,.
\ee
Using \rf{13042014-40},\rf{13042014-44},\rf{13042014-46},  we find the following relation:
\be \label{13042014-47}
G_{_{\pAdS_{d+1}}} |\xi_{_{\pAdS_{d+1}}} \rangle = \Uwh G_{_{\pAdS_{d+1}}}^\intm |\xi_{_{\pAdS_{d+1}}}^\intm \rangle\,,
\ee
where a new gauge transformation operator $G_{_{\pAdS_{d+1}}}$ appearing in \rf{13042014-47} takes the form
\beq
\label{13042014-48} G_{_{\pAdS_{d+1}}} & = &   \alphabf \Dbf - e_1 - \alphabf^2 \frac{1}{2N_\alphabf +d-1}\eb_1\,,
\\
\label{13042014-49} && e_1 =   \zeta \ewt_1 \chib \,, \qquad \eb_1 = - \chi  \ewt_1\bar\zeta\,,
\\
\label{13042014-50} && \ewt_1 \equiv \bigl(|\rho|(2s+d-5-2N_\zeta - N_\chi)\bigr)^{1/2} e_\zeta \,.
\eeq

Gauge transformation operator \rf{13042014-48}-\rf{13042014-50} coincides with the one we presented in \rf{12042014-04} in Sec.\ref{lagraarbspin}.   All that remains to get the Lagrangian of conformal field in Sec.\ref{lagraarbspin} is to find a realization of operator $m^2$ on space of $|\phi_{_{\pAdS_{d+1}}}\rangle$ in \rf{13042014-44}. To this end we should plug operators $e_1$, $\eb_1$ \rf{13042014-49} into the general formula for $m^2$ given in \rf{05042014-27}. Doing so, we get expression for $m^2$ given in \rf{04042014-28}.
Note that, for derivation of $m^2$ via the general formula in \rf{05042014-27}, one needs to use the second constraint in \rf{13042014-51}.

\bigskip

{\bf Acknowledgments}.

This work was supported by the RFBR Grant No.14-02-01172.

%%%%%%%%%%%%%%%%%%%%%%%%%%%%%%%%%%%%%%%%%%%%%%%%%%%%%%%%%%%%%%%%%%%%%%%%%
\setcounter{section}{0}\setcounter{subsection}{0}
\appendix{ Notation }
%%%%%%%%%%%%%%%%%%%%%%%%%%%%%%%%%%%%%%%%%%%%%%%%%%%%%%%%%%%%%%%%%%%%%%%%%

The vector indices of the Lorentz algebra $so(d,1)$ take the values $A,B,C,E=0,1,\ldots ,d$. We use the mostly positive flat metric tensor $\eta^{AB}$. To simplify
expressions we drop $\eta^{AB}$ in scalar products, i.e., we use the convention $X^AY^A
\equiv \eta_{AB}X^A Y^B$.

A covariant derivative $D^A$ is defined by the relations $D^A = \eta^{AB}D_B$,
\be \label{12042014-22}
D_A \equiv e_A^\mu D_\mu\,,  \hspace{0.5cm} D_\mu \equiv
\partial_\mu + \half \omega_\mu^{AB} M^{AB}\,, \hspace{0.5cm} M^{AB} = \alpha^A
\bar\alpha^B - \alpha^B \bar\alpha^A\,,
\ee
$\partial_\mu = \partial/\partial x^\mu$, where $x^\mu$ are the coordinates of $(A)dS_{d+1}$ space carrying the base manifold indices, $e_A^\mu$ is inverse vielbein
of $(A)dS_{d+1}$ space, $D_\mu$ is the Lorentz covariant derivative and the
base manifold index takes values $\mu = 0,1,\ldots, d$. The $\omega_\mu^{AB}$ stands for
the Lorentz connection of $(A)dS_{d+1}$ space, while $M^{AB}$ stands for a spin
operator of the Lorentz algebra $so(d,1)$. Fields in $(A)dS_{d+1}$ space carrying the flat indices, $\Phi^{A_1\ldots A_s}$, are related to contravariant tensor field, $\Phi^{\mu_1\ldots \mu_s}$, in a standard way, $\Phi^{A_1\ldots A_s} \equiv e_{\mu_1}^{A_1}\ldots e_{\mu_s}^{A_s} \Phi^{\mu_1\ldots \mu_s}$. The D'Alembert operator of (A)dS space is defined as
\be
\Box_{_\pAdS}   \equiv D^AD^A + \omega^{AAB}D^B\,, \qquad \omega^{ABC} \equiv e^{A \mu}\omega_\mu^{BC}\,,\qquad e \equiv \det e_\mu^A\,.
\ee

Operator $D_\mu$ given in \rf{12042014-22} is acting on the generating function constructed out of the oscillators $\alpha^A$. Using the notation $\DD_\mu$ for a realization of this operator on fields carrying flat indices we get
\be
\DD_\mu \phi^A = \partial_\mu \phi^A + \omega_\mu^{AB}(e)\phi^B\,,
\ee
Instead of $\DD_\mu$, we prefer to use a covariant derivative with the flat indices $\DD^A$, %
\beq
&& \DD_A \equiv e_A^\mu \DD_\mu\,,\qquad \DD^A = \eta^{AB}\DD_B\,,
\\
&& [\DD^A,\DD^B] \phi^C = R^{ABCE}\phi^E \,,
\eeq
where the Riemann tensor of (A)dS space is given by
\beq
\label{14042014-01} && R^{ABCE} = \rho (\eta^{AC}\eta^{BE} - \eta^{AE}\eta^{BC})\,,
\\[5pt]
\label{14042014-02} && \rho = \frac{\epsilon}{R^2}\,, \qquad \epsilon = \left\{ \begin{array}{cl} 1 & \hbox{for \ dS}
\\[5pt]
-1 & \hbox{for \ AdS}
\end{array}\right.
\eeq

For the Poincar\'e parametrization of $(A)dS_{d+1}$ space given in \rf{10042014-03}, the flat metric $\eta^{AB}$ takes the form as in \rf{12042014-24},\rf{12042014-25}, while  vielbein $e^A=e^A_\mu dx^\mu$ and Lorentz connection, $de^A+\omega^{AB}\wedge e^B=0$, are given by
\be \label{12042014-23}
e_\mu^A=\frac{R}{z}\delta^A_\mu\,,\qquad \omega^{AB}_\mu=\frac{1}{z}(\eta^{zA} \delta^B_\mu - \eta^{zB}\delta^A_\mu)\,.
\ee
When using the Poincar\'e parametrization, the coordinates of $(A)dS_{d+1}$ space
$x^\mu$ carrying the base manifold indices are identified with coordinates
$x^A$ carrying the flat vectors indices of the $so(d,1)$ algebra, i.e., we
assume $x^\mu = \delta_A^\mu x^A$, where $\delta_A^\mu$ is the Kronecker delta
symbol. With choice made in \rf{12042014-23}, the covariant derivative $D^A$ in \rf{12042014-22} takes the form $D^A= \frac{1}{R}(z \partial^A + M^{zA})$, $\partial^A=\eta^{AB}\partial_B$.

The Cartesian coordinates in $R^{d,1}$ are denoted by  $x^A$, while derivatives with respect to $x^A$ are denoted by $\partial_A$, $\partial_A \equiv
\partial/\partial x^A$.

Creation operators
$\alpha^A$, $\zeta$, $\upsilon^\oplussm$, $\upsilon^\ominussm$, $\vartheta$, $\chi$ and the respective annihilation operators
$\bar\alpha^A$, $\bar\zeta$, $\bar\upsilon^\oplussm$, $\bar\upsilon^\ominussm$, $\bar\vartheta$, $\bar\chi$  are referred to as oscillators.
Commutation relations of the oscillators, the vacuum $|0\rangle$, and
hermitian conjugation rules are defined as
\beq
\label{13042014-32} && \hspace{-0.7cm} [ \bar\alpha^A,\alpha^B] = \eta^{AB}, \quad [\bar\zeta,\zeta]=1,
\quad  [\bar\upsilon^\oplussm,\upsilon^\ominussm] = 1,\quad
[\bar\upsilon^\ominussm,\upsilon^\oplussm] = 1,
\quad [\varthetab,\vartheta] = -\epsilon,\quad [\chib,\chi] = \epsilon,\qquad
\\
&& \hspace{-0.6cm} \bar\alpha^A |0\rangle = 0\,,\hspace{1cm} \bar\zeta |0\rangle = 0\,,\hspace{1cm} \bar\upsilon^\oplussm |0\rangle = 0\,,\hspace{0.5cm} \bar\upsilon^\ominussm |0\rangle = 0\,,\qquad
\bar\vartheta |0\rangle = 0\,,\hspace{0.5cm} \bar\chi |0\rangle = 0\,,\qquad
\\
&& \hspace{-0.6cm} \alpha^{A \dagger} = \bar\alpha^A\,, \hspace{1cm} \zeta^\dagger = \bar\zeta\,,  \hspace{1cm} \upsilon^{\oplussm \dagger} = \bar\upsilon^\oplussm\,,  \hspace{1cm} \upsilon^{\ominussm\dagger} = \bar\upsilon^\ominussm\,, \hspace{0.8cm} \vartheta^\dagger = \varthetab\,,  \hspace{1cm} \chi^\dagger = \chib\,.
\eeq

The oscillators $\alpha^A$, $\bar\alpha^A$ and $\zeta$, $\bar\zeta$, $\upsilon^\oplussm$, $\bar\upsilon^\ominussm$,$\upsilon^\ominussm$, $\bar\upsilon^\oplussm$, $\vartheta$, $\bar\vartheta$, $\chi$, $\bar\chi$ transform in the respective vector and scalar representations of the Lorentz
algebra $so(d,1)$. Throughout this paper we use operators constructed out of the
derivatives and the oscillators,
\beq
&& \alphabf \Dbf \equiv \alpha^A D^A\,,\qquad \bar\alphabf \Dbf
\equiv \bar\alpha^A D^A\,,
\\
&& \alphabf \partialbf \equiv \alpha^A\partial^A\,,\qquad \bar\alphabf \partialbf
\equiv \bar\alpha^A \partial^A\,,
\\
&& \alphabf^2 \equiv \alpha^A \alpha^A\,,\qquad \ \bar\alphabf^2 \equiv
\bar\alpha^A \bar\alpha^A\,,
\\
&& N_\alphabf \equiv \alpha^A \bar\alpha^A  \,,
\qquad
N_\zeta \equiv \zeta \bar\zeta \,, \qquad N_\vartheta = -\epsilon \vartheta\varthetab\,, \qquad N_\chi \equiv \epsilon \chi\chib\,,
\\
&& N_{\upsilon^\oplussm} = \upsilon^\oplussm \bar\upsilon^\ominussm \,, \qquad  N_{\upsilon^\ominussm} = \upsilon^\ominussm \bar\upsilon^\oplussm\,,\qquad N_\upsilon = N_{\upsilon^\oplussm} + N_{\upsilon^\ominussm}\,,
\\
\label{13042014-15x} &&  \mubf \equiv 1 - \frac{1}{4} \alphabf^2\bar\alphabf^2\,,\hspace{1cm} \Pibf^\smponetwo \equiv 1 - \alphabf^2\frac{1}{2(2N_\alphabf + d+1)}\bar\alphabf^2\,,
\\
\label{13042014-15} && \Awt^A \equiv \alpha^A - \alphabf^2 \frac{1}{2N_\alphabf + d -1}\bar\alpha^A\,,
\\
\label{13042014-16} && e_\zeta \equiv \Bigl(\frac{2s+d-3-N_\zeta}{2s+d-3-2N_\zeta} \Bigr)^{1/2} \,.
\eeq

\end{document}